\documentclass[11pt,a4paper]{article}
\usepackage{amssymb}
\usepackage{amsmath}
\usepackage[dvips]{graphicx}
\usepackage[font=small,labelfont=bf]{caption}
\usepackage[]{bbm}

\newcommand{\hZ}{\hat{Z}}
\newcommand{\hM}{\hat{M}}
\newcommand{\be}{\begin{equation}}
\newcommand{\ee}{\end{equation}}
\newcommand{\Tr}{\textrm{Tr}}
\newcommand{\vc}[1]{\mbox{\boldmath $#1$}}
\newcommand{\dd}{\textrm{d}}
\newcommand{\undertilde}[1]{\underset{\widetilde{}}{#1}}
\newcommand{\wideundertilde}[1]{\underset{\widetilde{\hspace{1cm}}}{#1}}

\newcommand{\hlf}{\frac{1}{2}}
\newcommand{\Nc}{N_{\textrm{c}}}

\newcommand{\myRe}{\textrm{Re}}
\newcommand{\MSb}{\overline{\rm{MS}}}

\newcommand{\latint}[1]{\int_{-\frac{\pi}{a}}^{\frac{\pi}{a}}\frac{\dd^3{#1}}{(2\pi)^3}}
\newcommand{\Latint}[2]{\int_{-\frac{\pi}{a}}^{\frac{\pi}{a}}\frac{\dd^3{#1}}{(2\pi)^3}\frac{\dd^3{#2}}{(2\pi)^3}}
\newcommand{\prop}[2]{\frac{1}{\widetilde{{#1}}^2 + {#2}}}
\newcommand{\cont}{\mathcal{O}(a)}
\newcommand{\tadpoles}{\textrm{tadpoles.}}
\newcommand{\Det}{\textrm{Det}}

\begin{document}

\begin{titlepage}
\begin{flushright}
HIP-2007-18/TH\\
\end{flushright}
\begin{centering}
\vfill{}
 
{\Large{\bf Framework for non-perturbative analysis of a Z(3)-symmetric effective theory of finite temperature QCD}}

\vspace{0.8cm}

A. Kurkela\footnote{aleksi.kurkela@helsinki.fi}

\vspace{0.8cm}

{\em%
Theoretical Physics Division, 
Department of Physical Sciences, \\ 
P.O.Box 64, FI-00014 University of Helsinki, Finland\\ }

\vspace*{0.3cm}

\abstract{\noindent
We study a three dimensional Z(3)-symmetric effective theory of high temperature QCD. The exact lattice-continuum relations, needed in order to perform lattice simulations with physical parameters, are computed
to order $\mathcal{O}(a^0)$ in lattice perturbation theory.
Lattice simulations are performed to determine the phase structure of a subset of the parameter space.
}
\vfill
\noindent

\vspace*{1cm}
 
\noindent

\vfill
\end{centering}
\end{titlepage}

\section{Introduction}
At high temperature, QCD matter undergoes a deconfinement transition, where ordinary had\-ron\-ic matter transforms into strongly interacting quark-gluon plasma \cite{Adams:2005dq, Adcox:2004mh,Gyulassy:2004zy}. In the absence of quarks, $N_f=0$, the transition is a symmetry-breaking first order transition, where the order parameter is the thermal Wilson line \cite{McLerran:1981pb,Svetitsky:1982gs}. The non-zero expectation value of the Wilson line signals the breaking of the Z(3) center symmetry of quarkless QCD at high temperatures.

The transition has been studied extensively using lattice simulations \cite{Blum:1994zf,Karsch:2000ps,Heller:2006ub}. Thermodynamical quantities, condensates and various correlators can be measured on the lattice and the equation of state can be estimated. This approach, however, becomes computationally exceedingly expensive at high temperatures, and thus cannot be applied to temperatures $T$ above $\sim5T_c$.
The complementary approach has been to construct perturbatively effective theories, such as electrostatic QCD or EQCD, using the method of dimensional reduction \cite{Ginsparg:1980ef,Appelquist:1981vg,Kajantie:1995dw} to quantitatively describe high temperature regime of QCD \cite{Braaten:1995jr,Kajantie:2002wa,Hietanen:2006rc,Vepsalainen:2007ke}. In the dimensional reduction procedure, however, one expands the temporal gauge fields around one of the Z(3) vacua and thus explicitly violates the center symmetry. The range of validity of these theories therefore ends for $T$ below $\sim 5T_c$, where the fluctuations between different vacua become important. There have also been several attempts to the build models for Wilson line, respecting the center symmetry \cite{Stephanov:1991hs,Trappenberg:1992bt,Pisarski:2000eq,Sannino:2002wb,Bialas:2004gx,Pisarski:2006hz}. These models give a qualitative handle on the transition but cannot be perturbatively connected to QCD.

As a unification of these strategies, an effective field theory of high temperature QCD respecting the Z(3) center symmetry has been constructed in \cite{Vuorinen:2006nz}. At high temperatures, the effective theory reduces to EQCD guaranteeing the correct behavior there, but the model still preserves the center symmetry. The effective theory is further connected to full QCD by matching the domain wall profile separating two different Z(3) minima.

Being a three dimensional model, the new theory relies on the scale separation between the inverse correlation length and the lowest non-zero Matsubara mode, which is still modest at $T_c$ \cite{Arnold:1995bh,Philipsen:2001ip}. Thus, one hopes that the range of validity of this theory would extend down to $T_c$. The effective theory is a confining one, so perturbative analysis breaks down. Non-perturbative methods, i.e. lattice simulations, are thus needed to find out the physical properties, such as correlation lengths, condensates, and most importantly the phase structure of the theory, to test its regime of validity.

The effective theory is super-renormalizable, and thus the connection between the continuum $\MSb$ and lattice regulated theories can be obtained exactly to the desired order in the lattice spacing $a$. The matching of the parameters of the Lagrangian to order $\mathcal{O}(a^0)$, which is needed in order to perform simulations with $\MSb$ scheme parameters and to obtain physical results, requires a two-loop lattice perturbation theory calculation. The one-loop terms remove any linear $1/a$ divergences, while two-loop terms remove the logarithmic $\log(1/a)$ divergences and the constant differences in the mass terms of the theory. In addition, the condensates have lattice spacing dependence and constant differences between the two schemes and can be calculated to order $\mathcal{O}(a^0)$ by performing a two-loop calculation for operators up to cubic order and a four-loop calculation for the quartic condensates. In this paper we perform the needed two-loop calculations.

This paper is organized as follows. In Sections 2 and 3 we define the theory in continuum $\MSb$ regularization and on the lattice, respectively. In Section 4 we study the phase diagram of a subset of the parameter space of the theory. Details of the matching between the continuum and lattice theories are given in the appendices.

\section{Theory}
The theory we are studying is defined by a three dimensional continuum action, which we renormalize in the $\MSb$ scheme
\begin{equation}
S = \int \dd ^{3-2\epsilon}x  \left\{ \hlf \Tr F_{ij} ^2  + \Tr\left( D_i Z^\dagger D_i Z \right) + V_0(Z) + V_1(Z) \right\},\label{action}
\end{equation}
where
\begin{align}
F_{ij} &=\partial_i A_j - \partial_j A_i + i g_3 [A_i,A_j] \\
D_i &= \partial_i - i g_3 [A_i,\quad]
\end{align}
and $Z$ is a $3\times3$ complex matrix, which in the limit $\epsilon\rightarrow 0$ has dimension $\dim Z = \sqrt{\textrm{GeV}}$. The gauge fields $A_i$ are Hermitean traceless $3\times 3$ matrices and can be expressed using generators of SU(3), $A_i=A_i^a T^a$, with $\Tr T^aT^b=\hlf\delta^{ab}$ . The covariant derivative is in the adjoint representation. The potentials $V_0$, the ``hard'' potential, and $V_1$, the ``soft'' potential, are\footnote{Our notation is obtained from that in \cite{Vuorinen:2006nz} by scaling with $g_3$: $A_i \rightarrow g_3 A_i$, $Z\rightarrow g_3 Z$, $c_1 \rightarrow c_1$, $c_2 \rightarrow g_3^{-1} c_2$, $c_3 \rightarrow g_3^{-2} c_3$, $\tilde{c}_1 \rightarrow g_3^{-2}d_1$, $\tilde{c}_2 \rightarrow g_3^{-3} d_2$, and $\tilde{c}_3\rightarrow g_3^{-4} d_3$}:
\begin{align}
V_0(Z) &= c_1 \Tr[Z^\dagger Z] +2c_2 \myRe(\Det[Z])+c_3 \Tr[(Z^\dagger Z)^2], \label{V0}\\
V_1(Z) &=  d_1 \Tr[M^\dagger M] + 2 d_2 \myRe(\Tr[M^3]) + d_3 \Tr[(M^\dagger M)^2],\label{V1}
\end{align}
where $M=Z-\frac{1}{3}\Tr[Z] \mathbbm{1}$ is the traceless part of $Z$. Here, the gauge coupling $g_3$ has a positive mass dimension $\dim[g_3^2] =$GeV, making the theory
super-renormalizible. Because of the super-renormalizibility, the coefficients $c_2,c_3,d_2$, and $d_3$ are renormalization scale independent and only the mass terms $c_1$ and $d_1$ acquire a scale dependence in the $\MSb$ renormalization scheme. The scale dependence in the mass terms arises from a two-loop calculation and has the form:
\begin{align}
 c_1(\bar{\mu}) &= \frac{1}{16 \pi^2}\left[ 64 c_3 g_3^2 + \frac{88}{9}c_3^2\right] \log\left(\frac{\Lambda}{\bar{\mu}}\right)\label{msb_c1}\\
 d_1(\bar{\mu}) &= \frac{1}{16\pi^2}\left[ \frac{280}{9} c_3^2 - 64 d_3 g_3^2 + \frac{92}{3}\left(2 d_3 c_3 + d_3^2\right) + \frac{9}{2}g_3^4 \right] \log\left(\frac{\Lambda}{\bar{\mu}}\right),\label{msb_d1}
\end{align}
where $\Lambda$ is a constant specifying the theory and $\bar{\mu}$ is the $\MSb$ scale parameter. The coefficients $c_i$, $d_i$, and $g_3$ are matched to the parameters of full thermal QCD by imposing the condition that the theory reduces to EQCD at the high temperature limit, and that the theory reproduces the domain wall profile of full QCD \cite{Vuorinen:2006nz}. This defines a subset of parameter values (with a limited accuracy due to perturbative matching), for which the theory describes thermal QCD\footnote{In the matching, the hard potential is parametrically larger than the soft potential, explaining the terminology.}. However, in this paper we consider the model in general, and do not restrict ourselves only to the physical region. 

The action is defined only for the number of colors $\Nc=3$, but for generality, we give some of the perturbative results for any $\Nc$.
For analytic calculations the scalar field $Z$ can be expanded around the vacuum:
\be
Z= \sqrt{\frac{1}{6}} ( \phi + i \chi) \mathbbm{1} + (H + iA), \label{expansion}
\ee
where $\phi$ and $\chi$ are real scalars and $H$ and $A$ are Hermitean traceless matrices. Fields $H$ and $A$ can be written with
the generators of the SU(3) group, $H=H^aT^a$ and $A=A^aT^a$, where $H^a$ and $A^a$ are real scalars.

The action is invariant under local gauge transformations, with $Z$ transforming in the adjoint representation:
\begin{align}
A_i(\vc{x}) &\longrightarrow G(\vc{x}) \left(A_i(\vc{x}) -\frac{i}{g_3}\partial_i \right)G^{-1}(\vc{x}), \\
Z(\vc{x})  &\longrightarrow G(\vc{x}) Z(\vc{x}) G^{-1} (\vc{x}),
\end{align}
where $G(\vc{x})\in$SU(3). In addition to this, there are further global symmetries in the potentials. The potential $V_0$ is invariant in global SU(3)$\times$SU(3) transformations
\be
Z(\vc{x}) \longrightarrow L Z(\vc{x}) R,
\ee
where $L$ and $R$ are SU(3) matrices. The potential $V_1$ is invariant under Z(3) transformations
\be
M\rightarrow zM,
\ee
where $z=e^{i 2\pi n/3}$, which generalizes into a U(1) symmetry if $d_2=0$.
This implies that in the presence of the both potentials (with non-zero coefficients), the overall global symmetry of the Lagrangian is $Z\rightarrow zZ$.
\section{Lattice action}
In order to perform non-perturbative simulations, the theory has to be formulated on the lattice. On the lattice, the scalar field $Z$ lives on the sites of the lattice, and the gauge fields $A_i$ are traded for link variables $U_i$, which are elements of SU($\Nc$) and live on the links connecting adjacent sites.
The lattice action corresponding to the continuum theory can be written as $S = S_W + S_Z$, 
where 
\begin{align}
S_W = \beta &\sum_{x,i<j} \left[ 1 - \frac{1}{\Nc} \myRe \Tr[ U_{\mu\nu}] \right] 
\end{align}
is the standard the Wilson action with the lattice coupling constant $$\beta = \frac{2\Nc}{a g_3^2},$$ corresponding to a lattice spacing $a$.
The continuum limit is taken by $\beta \rightarrow \infty$, and there the Wilson action reduces to the ordinary pure gauge action. 

The kinetic term, $\Tr\left( D_i Z^\dagger D_i Z \right)$, is discretized by replacing the covariant derivatives by covariant lattice differences. Then the scalar sector of the action reads: 
\begin{align}
 S_Z = 2\left(\frac{  2\Nc }{ \beta } \right)& \sum_{x,i}\myRe \Tr \left[\hZ^\dagger\hZ - \hZ^\dagger(x)U_i(x)\hZ(x+\hat{i})U^\dagger_i(x)\right] \nonumber\\
+ \left(\frac{2 N_c }{ \beta}\right) ^3 & \sum_x \left( \hat{c}_1\Tr[\hZ^\dagger \hZ] + 2\hat{c}_2 \myRe\Det{\hZ} + \hat{c}_3 \Tr[(\hZ^\dagger \hZ)^2] \right)\nonumber \\
+ \left(\frac{2 N_c }{ \beta} \right)^3 & \sum_x \left( \hat{d}_1\Tr[\hM^\dagger \hM] + 2\hat{d}_2 \myRe\Tr{\hM^3} + \hat{d}_3 \Tr[(\hM^\dagger \hM)^2] \right).
\end{align}
where $\hat{c}_i, \hat{d}_i$ are dimensionless numbers, and $\hM$ and $\hZ$ are dimensionless $\Nc \times \Nc$ complex matrices. Only the mass terms $\hat{c}_1$ and $\hat{d}_1$ require non-trivial renormalization and 
all the other terms can be matched to order $\mathcal{O}(a^0)$ on tree-level by simply scaling with $g_3$:
\begin{eqnarray}
Z  =g_3 \hZ,         & M   = g_3 \hM \\
c_2=g_3^3 \hat{c}_2, & d_2 = g_3^3 \hat{d}_2\\
c_3=g_3^2 \hat{c}_3,       & d_3 = g_3^2\hat{d}_3.
\end{eqnarray}
For the mass terms, renormalization has to be carried out, so that the physical masses of the fields are the same in both regularization schemes. A two-loop calculation gives (the details of the calculation and the definitions of the numerical constants are given in the appendix):
\begin{align}
\hat{c}_1 =& \frac{c_1}{g_3^4}  -  \frac{\Sigma}{4\pi} 2 \hat{c}_3 \beta - \frac{1}{16\pi^2} \left[\left(64 \hat{c}_3 + \frac{88}{9} \hat{c}_3^2 \right) \left( \log \beta +\zeta \right)  +\left[16\Sigma^2-64 \delta \right]  \hat{c}_3 \right]  + \mathcal{O}(\beta^{-1}) \nonumber \\
	=& \frac{c_1}{g_3^4} - \frac{1}{4 \pi} 6.3518228 \hat{c}_3 \beta \nonumber\\
 & \quad -\frac{1}{16\pi^2}\left[\left(64 \hat{c}_3 + \frac{88}{9} \hat{c}_3^2 \right) \left( \log \beta + 0.08849 \right)  + 37.0863  \hat{c}_3 \right]  + \mathcal{O}(\beta^{-1}) 
\label{c1_ren}
\end{align}
and
\begin{align}
\hat{d}_1 =& \frac{d_1}{g_3^4} - \frac{\Sigma}{4\pi} \beta (1+\frac{16}{9}\hat{d}_3) 
\nonumber \\ &+ \frac{1}{16\pi^2 } \Big\{ - \left(16 \Sigma^2-64 \delta\right)\hat{d}_3
\nonumber \\ &+ \left( \frac{280}{9} \hat{c}_3^2 - 64 \hat{d}_3 + \frac{92}{3}\left(2 \hat{d}_3 \hat{c}_3 + \hat{d}_3^2\right) + \frac{9}{2} \right) \left[\log\beta + \zeta\right] 
\nonumber \\ & - 9 \left[ \frac{5}{8}\Sigma^2 + \frac{19}{54}\pi\Sigma - 4\delta - 6\rho + 2 \kappa_1 - \kappa_4 \right]  \Big\} + \mathcal{O}(\beta^{-1})  \nonumber \\
 =& \frac{d_1}{g_3^4} - \frac{\beta}{4 \pi} \left(3.17591 + 5.64606  \hat{d}_3 \right) \nonumber\\
 -& \frac{1}{16 \pi^2} \Big\{41.780852 +37.0863 \hat{d}_3 \nonumber\\
-& \left( \frac{280}{9} \hat{c}_3^2 - 64 \hat{d}_3 + \frac{184}{3}\hat{d}_3 \hat{c}_3 + \frac{92}{3}\hat{d}_3^2 + \frac{9}{2}\right) \left[\log\beta + 0.08849\right] \Big\}+\mathcal{O}(\beta^{-1}). \label{d1_ren}
\end{align}
Here, we have set the renormalization scale to be $\bar{\mu}=g_3^2$ in Eqs. (\ref{msb_c1}) and (\ref{msb_d1}), and denote $c_1=c_1(g_3^2)$ and $d_1=d_1(g_3^2)$. By making this choice, we get the logarithmic term to be a function of the lattice coupling constant $\beta$. There are also higher order corrections (corrections of order $\mathcal{O}(\beta^{-1})$ corresponding to order $\mathcal{O}(a)$ in lattice spacing), but their effect vanishes in the continuum limit. 

Various operators also need to be renormalized on the lattice in order to convert their expectation values to continuum regularization. The $Z_3$-symmetry protects the lowest dimensional condensate $\langle g_3^{-1}\Tr Z\rangle$ from acquiring any additive renormalization, while a two-loop calculation gives for the quadratic condensates:
\begin{align}
 \langle g_3^{-2}\Tr Z^\dagger Z \rangle_{\MSb} &= \langle \Tr \hat{Z}^\dagger \hat{Z} \rangle_a -\left[\Nc\frac{\Sigma}{8\pi}\beta + \frac{1}{16\pi^2} 2\Nc(\Nc^2-1)\left(\log\beta + \zeta + \frac{\Sigma^2}{4}-\delta\right) \right] + \mathcal{O}(\beta^{-1})  \nonumber\\
&=\langle \Tr \hat{Z}^\dagger \hat{Z} \rangle_a-\left[0.3791 \beta + 0.3040(\log \beta + 0.66796) \right]+ \mathcal{O}(\beta^{-1}) \label{quadratic_condensates},\\
\langle g_3^{-2}\Tr M^\dagger M \rangle_{\MSb} &= \langle \Tr \hat{M}^\dagger \hat{M} \rangle_a -\bigg[ \frac{\Nc^2-1}{\Nc}\frac{\Sigma}{8\pi}\beta \nonumber\\ &\qquad\qquad+ \frac{1}{16\pi^2} 2\Nc(\Nc^2-1)\left(\log\beta + \zeta + \frac{\Sigma^2}{4}-\delta\right) \bigg] + \mathcal{O}(\beta^{-1}),\\
\langle g_3^{-2} \Tr Z^\dagger \Tr Z \rangle_{\MSb} &= \langle \Tr \hat{Z} ^\dagger \Tr \hat{Z}\rangle_a - \frac{3}{\Nc} \frac{\Sigma}{8\pi} \beta+ \mathcal{O}(\beta^{-1}),
\end{align}
where the subscript $a$ denotes the lattice regularization. For the cubic condensates we get:
\begin{align}
\langle g_3^{-3}2\myRe\Det Z \rangle_{\MSb} &= \langle 2\myRe\Det \hat{Z} \rangle_{a} \nonumber\\
&\hspace{-1.5cm}-\frac{1}{16\pi^2}\left[ \left( \frac{\Nc^2-1}{3}+\frac{4}{9}\right) \hat{c}_2 +  \left(8/\Nc-10\Nc+2\Nc^3 \right)(\frac{1}{3}\hat{c}_2+\hat{d}_2) \right]\left(\log\beta + \zeta\right)+\mathcal{O}(\beta^{-1})\\
\langle g_3^{-3}2\myRe\Tr M^3\rangle_{\MSb} &= \langle 2 \myRe \Tr \hat{M}^3 \rangle_{a}-\frac{1}{16\pi^2}\left[\frac{24}{\Nc}-30\Nc + 6\Nc^3\right](\frac{1}{3}\hat{c}_2 + \hat{d}_2)\left(\log\beta + \zeta\right)+\mathcal{O}(\beta^{-1})
\end{align}

The effect of subtraction of the divergences can be seen in Fig.1. The renormalization of the quartic operators to order $\mathcal{O}(\beta^0)$ would require a four-loop calculation, which we do not perform here since they are not measured at this stage.

\begin{figure}[ht]
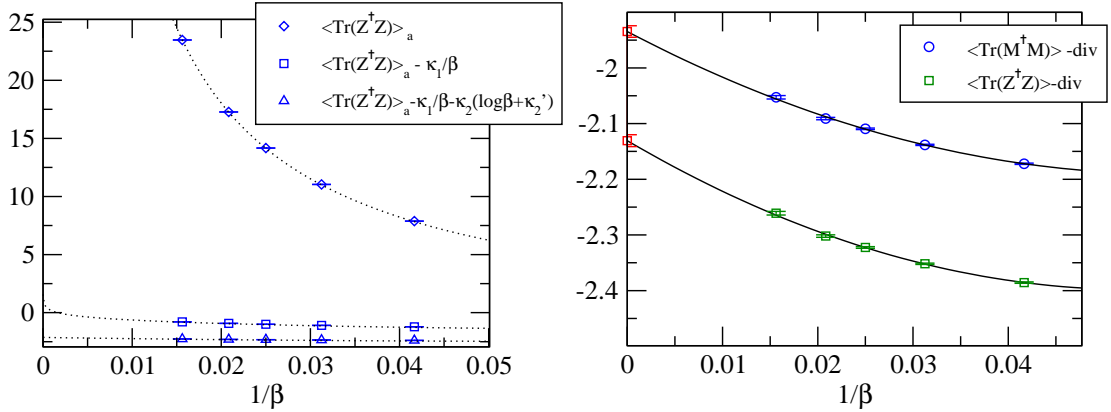

\begin{center}
\includegraphics*[width = 0.46\textwidth]{figure1a.eps}
\includegraphics*[width = 0.44\textwidth]{figure1b.eps}
\caption{The effect of subtraction of the divergences in $\langle \Tr Z^\dagger Z \rangle$ and $\langle \Tr M^\dagger M \rangle$ in a fixed physical volume with $d_1=6.6$ and $d_3=0.01$. On the left panel, the effect of subtracting the divergent parts of $\langle \Tr Z^\dagger Z \rangle$ is plotted. The constants $\kappa_1$, $\kappa_2$ and $\kappa_2'$ are the coefficients of the linear, logarithmic, and constant differences between lattice and $\MSb$ regularizations form equation (\ref{quadratic_condensates}). On the right panel: the continuum limit of the condensates.  Notice the negative values of the quadratic condensates in the symmetric phase.}\label{contlimit}\end{center}
\end{figure}

\section{Phase diagram of the soft potential}
A simpler model is obtained from the original theory by setting $c_i=0$. In this model, the trace of $Z$ decouples and can be integrated over as a free scalar field. The relevant degree of freedom is thus a traceless complex matrix $M$, or two traceless hermitian matrices $H$ and $A$. This can be viewed as a natural generalization\footnote{In EQCD with gauge group SU(3), there is only one linearly independent quartic gauge invariant operator namely $\Tr A_0^4$. In the complex case, however, there are four different $Z_3$-symmetric operators: $\Tr(M^\dagger M)^2$, $(\Tr M^\dagger M)^2$, $\Tr[M^\dagger M^\dagger M M]$ and $\Tr[{M^\dagger}^2]\Tr[M^2]$. In the case of unitary $M$, i.e. in the minimum of the hard potential, these operators collapse into a single one. However, since there is no such restriction in our model, the operators are linearly independent.   From these operators, we choose to include only the one appearing in the original theory, $\Tr( M^\dagger M)^2$.} of EQCD to complex values of the adjoint higgs field $A_0^a$.

The simpler model is defined by the action:
\begin{align}
S &= \int \dd^3x \left[\frac{1}{2}\Tr F_{ij}^2+ \Tr D_i M^\dagger D_i M + d_1 \Tr M^\dagger M+2d_2 \myRe(\Tr[M^3])+d_3 \Tr( M^\dagger M)^2\right] \\
&= \int \dd^3x \Big\{\frac{1}{2}\Tr F_{ij}^2 + \Tr D_i A D_i A + \Tr D_i H D_i H + d_1 \Tr A^2 + d_1 \Tr H^2 \nonumber \\&\quad+ 2 d_2 \Tr[H^3-3H A^2] + d_3 \Tr[H^4 + A^4 + 4H^2A^2-2HAHA] \Big\}.
\end{align}

If the cubic term $d_2$ is zero, the Lagrangian is invariant under a U(1) global symmetry $M \rightarrow gM$, $g \in $U(1). The breaking of the symmetry is signalled by a local order parameter:
\be
	\mathcal{A}=\sqrt{\langle\Tr A^3\rangle^2 + \langle\Tr H^3\rangle^2}.
\ee
This operator remains a valid order parameter after the renormalization since it has no additive renormalization, if $d_2=0$. In the symmetric phase $\mathcal{A}$ is strictly zero and in the broken phase the order parameter obtains a non-zero vacuum expectation value, while the two phases are separated by a first order transition. In the broken phase $\langle \Tr M^\dagger M\rangle$ is larger than in the symmetric phase.
After the inclusion of the cubic term, $\mathcal{A}$ is no longer strictly an order parameter, since the U(1) symmetry is explicitly broken. However, the first order transition remains and is accompanied with a significant discontinuity in $\mathcal{A}$ and $\langle \Tr M^\dagger M \rangle$.

\subsection{Perturbation theory}
In the limit of small $d_3/g_3^2$, the transition becomes very strong, and we expect a semiclassical approximation to produce the correct behavior of the critical line \cite{Kajantie:1998yc,Bronoff:1998hr}. We parametrize a constant diagonal hermitian background field in a fixed Landau gauge as follows:
\be
\langle M \rangle = 2pT_3+2\sqrt{3} q T_8 = 
\left(\begin{array}{ccc} 
q + p & 0& 0\\
0&q-p & 0\\
0&0&-2q
\end{array}\right),
\ee
where $p$ and $q$ are real scalars with dimensions of $g_3$.

Lattice simulations suggest that the $A\rightarrow-A$ symmetry is not broken spontaneously at any non-zero value of $d_2$, so that it is sufficient to consider only hermitian background fields. Using this parametrization, the 1-loop effective potential $V_1(d_1,d_2,d_3;p,q)$ can be calculated:

\begin{align}
 V_1(d_1&,d_2,d_3;q,p)=
(2p^2+6 q^2) d_1+36 q(p^2-q^2) d_2+\hlf(2p^2+6 q^2)^2 d_3\nonumber\\
-\frac{1}{3 \pi }&(8 |p|^3+|p-3q|^3+|p+3 q|^3) g_3^3\nonumber\\
-\frac{1}{12 \pi }&\Big\{2 \left[d_1+3 (p-q) d_2+2 (p^2+3 q^2) d_3\right]^{3/2}\nonumber\\
+&2 \left[d_1+6 q d_2+2 (p^2+3 q^2) d_3\right]^{3/2}\nonumber\\
+&2 \left[d_1-3 (p+q) d_2+2 (p^2+3 q^2) d_3\right]^{3/2}\nonumber\\
+&\left[d_1+4 (p^2+3 q^2) d_3-2 \sqrt{3 (p^2+3 q^2) d_2^2+18 q(p^2-q^2) d_2 d_3+(p^2+3 q^2)^2 d_3^2}\right]^{3/2}\nonumber\\
+&\left[d_1+4 (p^2+3 q^2) d_3+2 \sqrt{3 (p^2+3 q^2) d_2^2+18 q(p^2-q^2) d_2 d_3+(p^2+3 q^2)^2 d_3^2}\right]^{3/2}\Big\}\nonumber\\
-\frac{1}{12 \pi }&\Big\{2 \left[d_1-6 q d_2+2 (3 p^2+q^2) d_3\right]^{3/2}\nonumber\\
+&2 \left[ d_1+3 (p+q) d_2+2 (p^2-4 p q+7 q^2) d_3\right]^{3/2}\nonumber\\
+&2 \left[d_1-3 (p-q) d_2+2 (p^2+4 p q+7 q^2) d_3\right]^{3/2}\nonumber\\
+&\left[d_1+\frac{4}{3} (p^2+3 q^2) d_3-\frac{2}{3} \sqrt{27 (p^2+3 q^2) d_2^2+54 q (q^2-p^2) d_2 d_3+(p^2+3 q^2)^2 d_3^2}\right]^{3/2}\nonumber\\
+&\left[d_1+\frac{4}{3} (p^2+3 q^2) d_3+\frac{2}{3} \sqrt{27 (p^2+3 q^2) d_2^2+54 q (q^2-p^2) d_2 d_3+(p^2+3 q^2)^2 d_3^2}\right]^{3/2}\Big\},\label{effpot}
\end{align}
where the first term is the classical potential, the second one comes from one-loop vector diagrams and the fourth and the fifth from one-loop scalar diagrams of $H$ and $A$, respectively\footnote{By dropping the last term, i.e., the five last lines and scaling $d_1\rightarrow y$, $d_2\rightarrow i\gamma_3$, and $d_3 \rightarrow 2x$, one obtains the effective potential for EQCD in the presence of a finite (imaginary) chemical potential using the notation of \cite{Hart:2000ef}}. The effective potential has a symmetry arising from the permutations of the diagonal elements of the background $\langle M \rangle$ and has the following invariance:
\be
V_1(d_1,d_2,d_3;q,p=\pm 3q)=V_1(d_1,d_2,d_3;-2q,0).
\ee
More generally, the potential is invariant under rotations of $2\pi/3$ in the $(p,\frac{3}{2}q)$-plane and in the reflections of $p$:
\be
\begin{array}{ccccc}
\left( \begin{array}{rcl}
 q&\rightarrow&\frac{p-q}{2}\\
p&\rightarrow& \frac{p+3q}{2} 
\end{array}\right)
&
,
&
\left( \begin{array}{rcl}
q&\rightarrow&\frac{-p-q}{2}\\
p&\rightarrow&\frac{p-3q}{2}
\end{array}\right)
&
\textrm{and}
&
\left( \begin{array}{rcl}
q&\rightarrow&q\\
p&\rightarrow&-p
\end{array}\right)
\end{array}
\ee
Thus there is a fundamental region, which determines the potential over the whole plane. We choose the fundamental region to be bounded by the two lines $p=0$ and $p=-3q$ together with the condition $p\geq0$.
\begin{figure}[ht]
\begin{center}
\includegraphics*[width = 7cm]{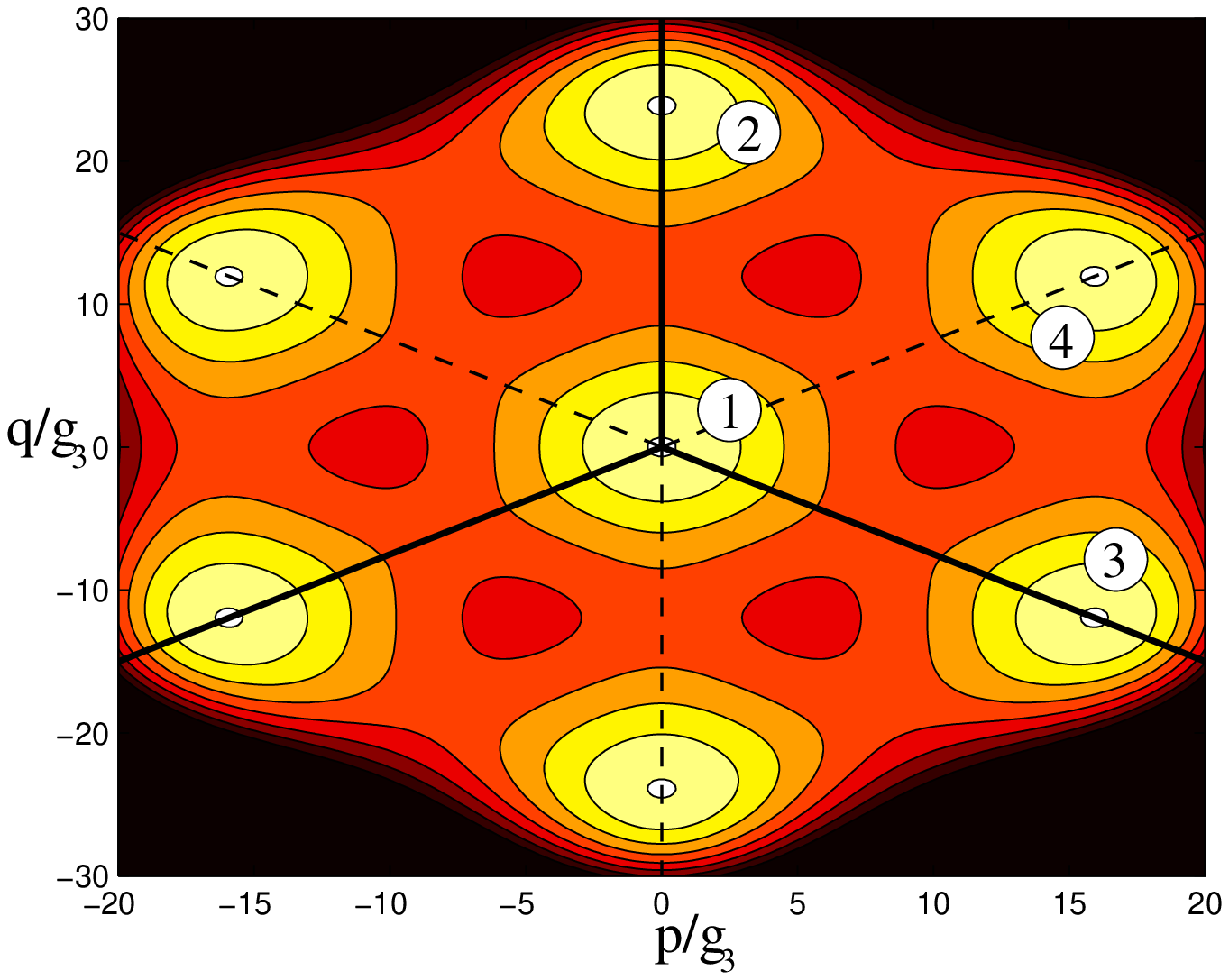}
\includegraphics*[width = 7cm]{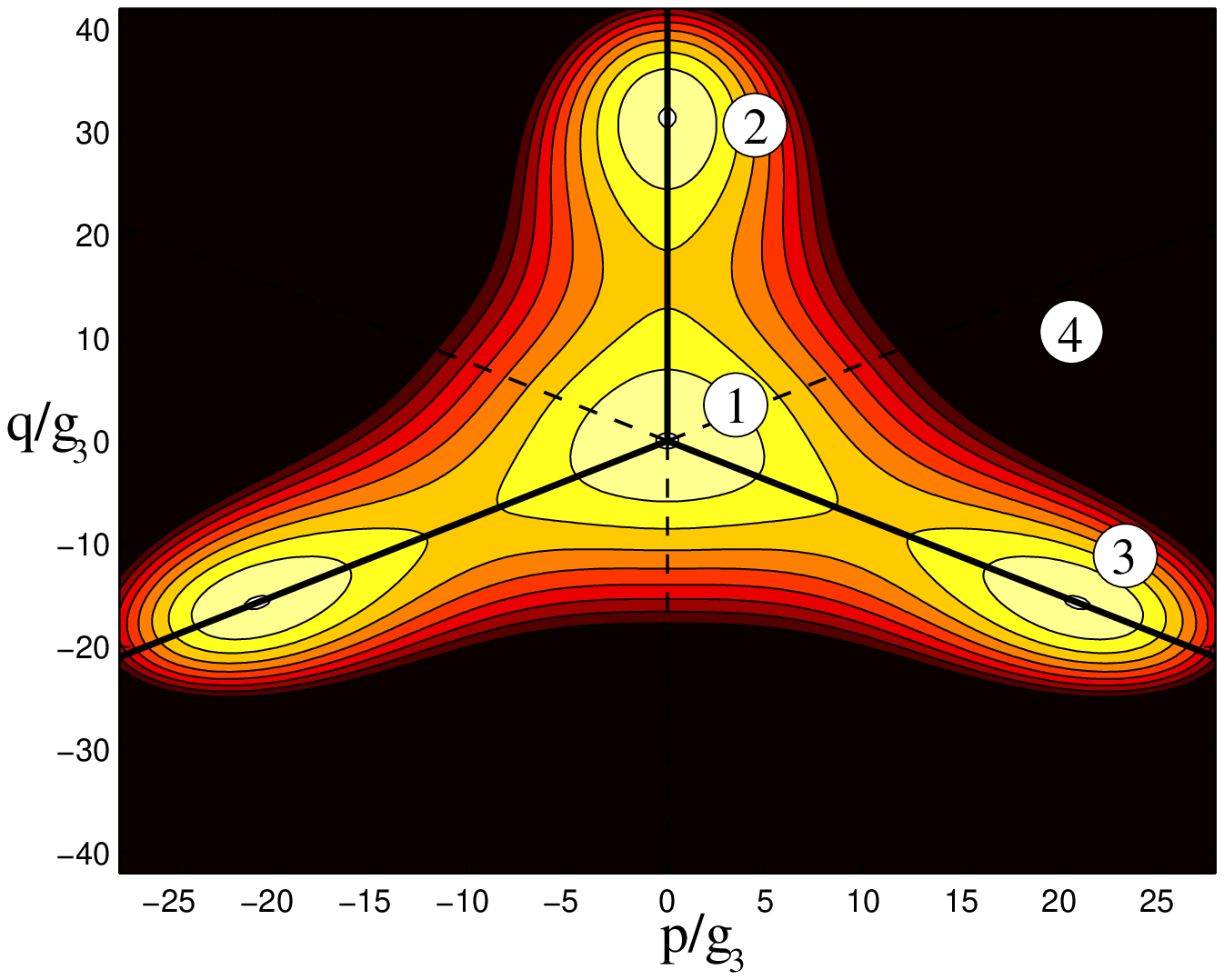}
\caption{1-loop effective potential in the ($q$,$p$)-plane at the critical point for $d_3=0.01$ and $d_2=0$ (left panel) and $d_2=0.05$ (right panel). Light areas represent the minima of the potential. Solid lines separate the three identical sectors which are related by the permutation symmetry of the diagonal elements of the background field $\langle M \rangle$. In the absence of $d_2$ there is an additional $U(1)$ symmetry making the directions marked with dashed lines identical to the $p=0$ direction. This symmetry explicitly broken by finite $d_2$ as seen on the right panel. }\label{effpotfig}\end{center}
\end{figure}
\begin{figure}[ht]
\begin{center}
\includegraphics*[width = 0.6\textwidth]{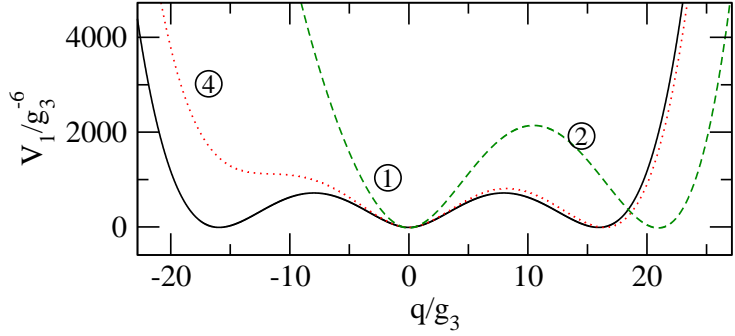}
\caption{1-loop effective potential with $p=0$ as a function of $q$ at the critical point for $d_3=0.01$ and $d_2=0$(line), 0.005(dotted),and 0.05(dashed). }\label{effpotmin}\end{center}
\end{figure}

In the fundamental region, there can be four different minima at the critical parameter values $d_1$, $d_2$ and $d_3$. The one at the origin (denoted by 1 in fig.\ref{effpotfig}) is the symmetric minimum, the minima 2 and 3 are connected by the permutation symmetry and correspond to the same physical broken minimum, with $\Tr H^3<0$ and $\Tr A^3=0$. The minimum 4 corresponds to a phase with $\Tr H^3>0$ and $\Tr A^3 = 0$ and is connected continuously to the minimum 2 by a global $U(1)$ symmetry if $d_2=0$. If $d_2\neq0$, the U(1) symmetry is lost and the minima 2 and 4 are no longer equivalent. If $d_2>0$ the minimum at 2 is favored over 4 and vice versa.

\begin{figure}[ht]
\begin{center}
\includegraphics*[width = 0.8\textwidth]{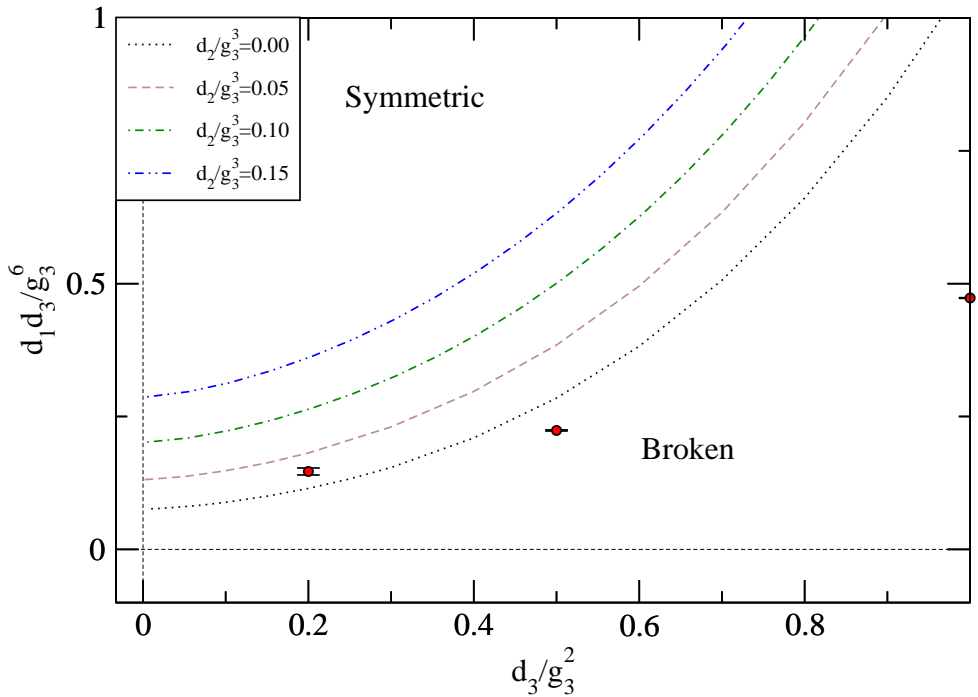}
\caption{1-loop perturbative phase diagram of the soft potential, $V_1$, as function of $d_1,d_2$ and $d_3$. A first order critical line separates the two phases. The symmetric phase refers to the phase where with $d_2=0$ the order parameter vanishes and with $d_2\neq0$ is smaller than in the broken phase. The data points represent non-perturbative lattice measurements with $d_2=0$ on a $N^3=12^3$ lattice.
The perturbative result approaches the lattice data points for small values of $d_3$ where the transition is very strong. Small discrepancy between the perturbative result and lattice data points at small $d_3/g_3^2$ is mostly due to finite volume effects.}\label{phasediagram6}\end{center}
\end{figure}

\begin{figure}[ht]
\begin{center}
\includegraphics*[width = 8.6cm]{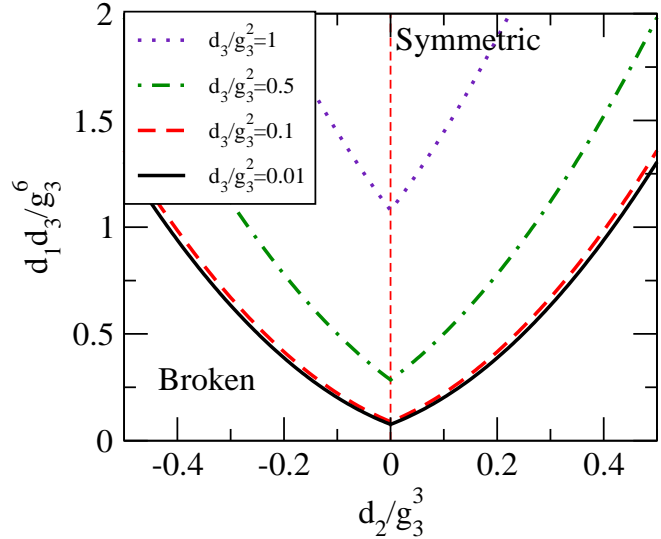}
\caption{1-loop perturbative phase diagram of the soft potential, $V_1$, as function of $d_1$ and $d_2$ with $d_3=2$. A first order critical line separates the two phases. The non-analyticity at $d_2=0$ is due to the change of global minimum between minima 3 and 4.}\label{phasediag_d2_pert}\end{center}
\end{figure}

Setting $d_2$ to zero and expanding in $d_3$ up to order $\mathcal{O}(d_3^2)$ the potential reads (for $p=0$):
\be
V_1(d_1,0,d_3;q,0)=18 q^2 d_3 \left[\left(|q|-\frac{g_3^3}{2\pi d_3}\right)^2+\frac{1}{3d_3}\left(d_1-\frac{3 g_3^6}{4\pi^2d_3}-\frac{24\sqrt{d_1}d_3}{9\pi}\right)\right]+\frac{4 d_1^{3/2}}{3\pi}+\mathcal{O}(d_3^2),
\ee
so that in the limit $d_3\rightarrow 0$ the potential has two coexisting minima and a first order transition for
\be
d_1=d_1^\textrm{crit}=\frac{3 g_3^6}{4\pi^2 d_3} \approx 0.0759909\frac{g_3^6}{d_3},
\ee
This sets the scaling of the critical line as a function of $d_3$ at small $d_3$. Corrections to this, and $d_2$ dependence, are obtained by minimizing the real part of Eq.(\ref{effpot}) numerically. The results are shown in Figs.\ref{phasediagram6} and \ref{phasediag_d2_pert}.
The phase transition is accompanied with a discontinuity in $q$:
\be
\Delta q = |q_\textrm{broken}-q_\textrm{symmetric}|= \frac{g_3^3}{2\pi d_3},\label{discont}
\ee
We see that the transition gets stronger as the coupling $d_3$ gets smaller justifying a posteriori the semiclassical approximation.

\subsection{Lattice analysis}
The perturbative calculation is valid only for small $d_3$ and a non-perturbative lattice analysis has to be performed to obtain the full phase structure of the model.  For the simulations we used a hybrid Monte-Carlo algorithm for the scalar fields and Kennedy-Pendleton quasi heat bath and full group overrelaxation for the link variables \cite{Duane:1987de,Kennedy:1985nu,deForcrand:2005xr}. 

The transition was found to be of the first order for all parameter values used in the simulations  ($d_3\leq4$ and $d_2\leq0.15$) accompanied with a large latent heat and surface tension; hysteresis curves showing discontinuity around critical point in $\langle \Tr M^\dagger M \rangle_{\MSb}$ can be seen in Fig.\ref{hyster}. The probability distributions of $\Tr M^\dagger M$ along the critical curve are very strongly separated (see Fig.\ref{histgram}). This makes the system change its phase very infrequently during a simulation, and multicanonical algorithm is needed to accommodate a phase flip in reasonable times for any system of a modest size \cite{Berg:1992qu}. Even with the multicanonical algorithm, the critical slowing restricts us to physical volumes up to $V\lesssim 50/g_3^6$. 

\begin{figure}[ht]
\begin{center}
\includegraphics*[width = 8.6cm]{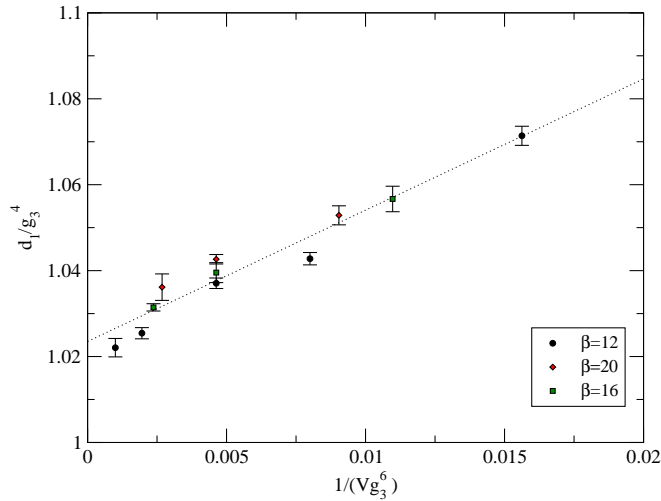}
\caption{Volume dependence of the pseudo-critical point with $d_3=2$ and $d_2=0.1$. The pseudo-critical point was determined by requiring equal probability weight for $\Tr M^\dagger M$ in both phases. The line represents a linear fit. The dependence on lattice spacing and volume seem to be within 5\% for the lattice spacings and volumes used. }\label{finite_size}\end{center}
\end{figure}

\begin{figure}[ht]
\begin{center}
\includegraphics*[width = 0.95\textwidth]{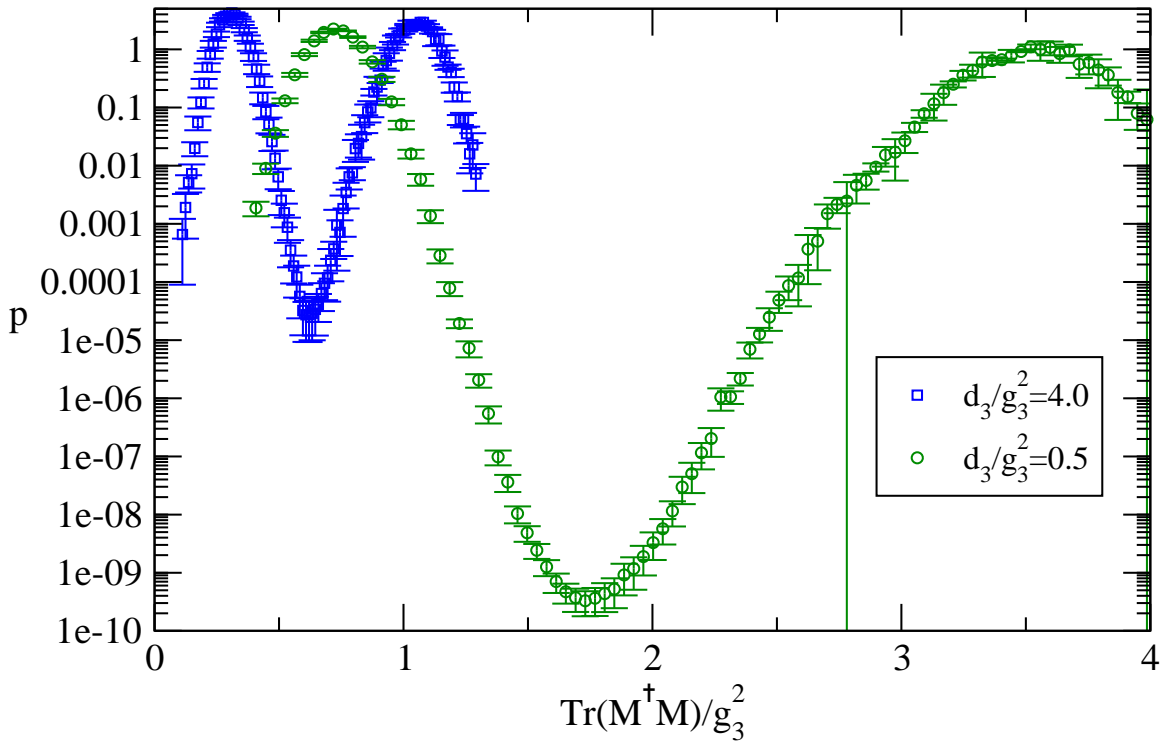}
\caption{Histograms of $\Tr M^\dagger M$ in logarithmic scale with $d_2=0$ along the critical curve. Transition channel between the peaks weakens and the transition gets stronger for decreasing $d_3$. For $d_3/g_3^2=0.5$, the relative probability density in the tunneling channel is suppressed by a factor $\sim10^{-10}$.}\label{histgram}\end{center}
\end{figure}

\begin{figure}[ht]
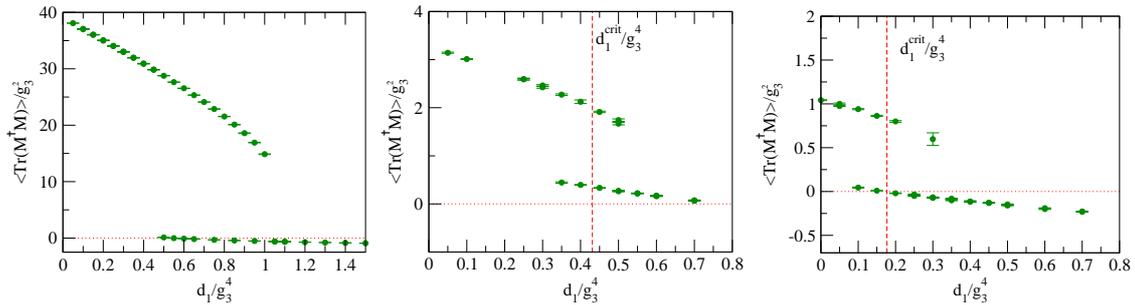

\begin{center}
\includegraphics*[width = 0.3\textwidth]{figure8a.eps}
\includegraphics*[width = 0.3\textwidth]{figure8b.eps}
\includegraphics*[width = 0.31\textwidth]{figure8c.eps}
\caption{Discontinuity in the quadratic condensate in continuum regularization $\langle\Tr M^\dagger M\rangle_{\MSb}$ for $d_3=0.1,1,3$. The phase transition gets weaker as the coupling $d_3$ grows. The metastable regions shrink and the discontinuity diminishes.}\label{hyster}\end{center}
\end{figure}

The pseudo-critical point was determined requiring equal probability weight for $\Tr M^\dagger M$ in both phases.  The simulations were performed with $\beta=12$ and a lattice size $N^3=12^3$, which precludes the continuum extrapolation as well as the thermodynamical limit. However, these limits were studied for one set of parameter values and the dependence of the critical point on both lattice spacing and volume were found to be of order of five per cent for the lattice spacings and volumes used (see Fig.\ref{finite_size} and Table \ref{lattices}).

\begin{figure}[ht]
\begin{center}
\includegraphics*[width = 0.8\textwidth]{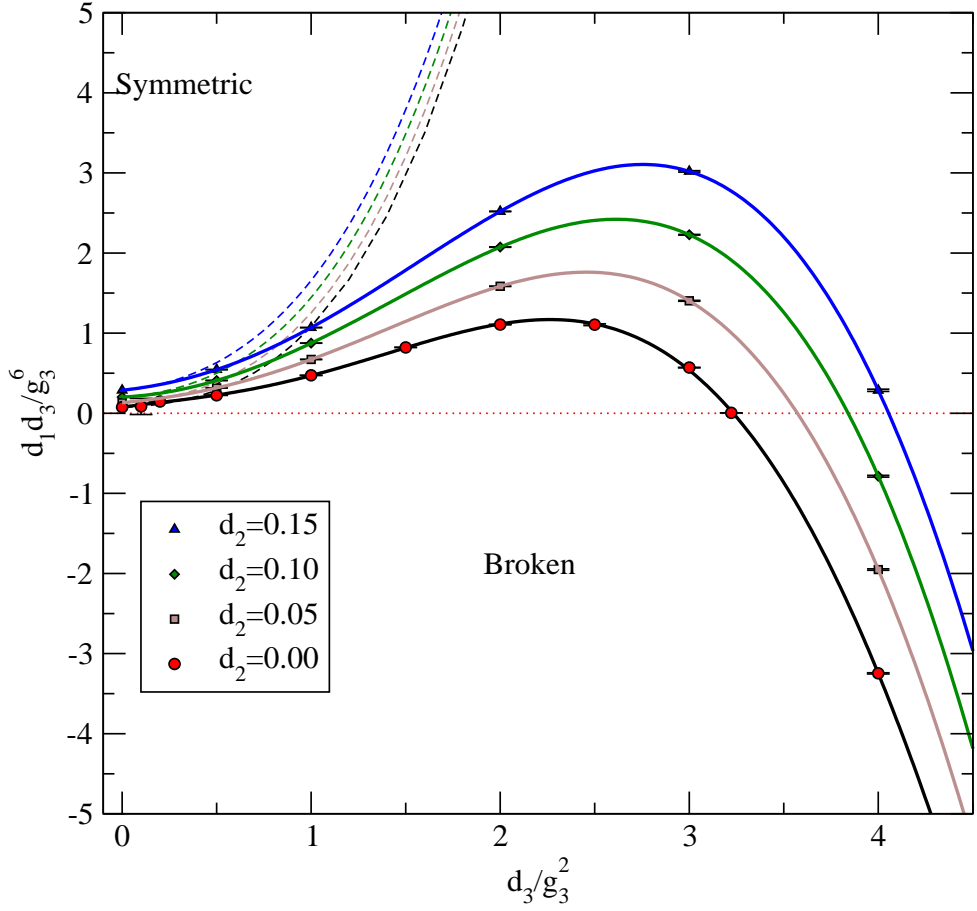}
\caption{The phase diagram of the soft potential as a function of $d_1,d_2$ and $d_3$. First order critical line separates two phases. Solid lines represent polynomial fits to the lattice data points and dashed lines are the perturbative predictions. The symmetric phase refers to the phase where with $d_2=0$ the order parameter $\mathcal{A}$ vanishes and with $d_2\neq0$ is smaller than in the broken phase. Also $\langle \Tr M^\dagger M \rangle$ is significantly smaller in the symmetric phase and the critical line was determined requiring equal propability weight for $\langle \Tr M^\dagger M \rangle$ in both phases. }\label{phasediagram5}\end{center}
\end{figure}

\begin{figure}[ht]
\begin{center}
\includegraphics*[width = 0.8\textwidth]{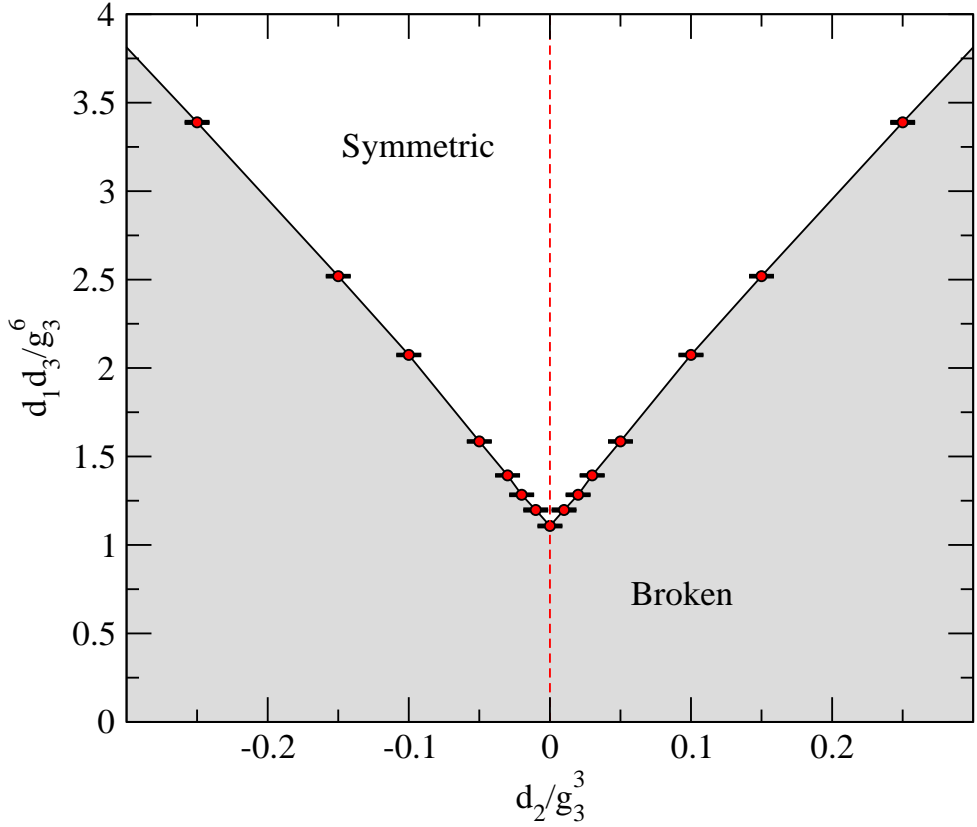}
\caption{The the phase diagram as a function of $d_2$, with $d_3=2$. The symmetric phase refers to the phase where with $d_2=0$ the order parameter vanishes and with $d_2\neq0$ is smaller than in the broken phase. }\label{phasediag_d3_2}\end{center}
\end{figure}

\begin{figure}[ht]
\begin{center}
\includegraphics*[width = 10cm]{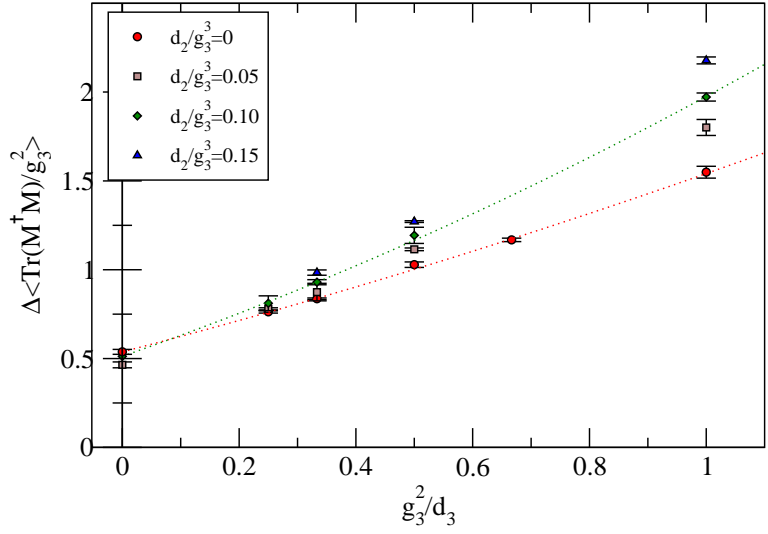}
\caption{Discontinuity in $\langle \Tr M^\dagger M \rangle$ along the critical line $d_1^\textrm{crit}(d_2,d_3)$. Dotted lines represent second order polynomial fits to the data, and the points on the y-axis represent extrapolations to infinite $d_3$. The large $d_3$ extrapolation yields a finite value suggesting that the transition remains of first order even at large $d_3$.
 }\label{latent}\end{center}
\end{figure}

The phase diagram can be seen in Fig.\ref{phasediagram5} and Fig.\ref{phasediag_d3_2}. The non-perturbative critical line follows the perturbative one for small values of $d_3$, but for larger $d_3$ fluctuations make the system prefer the symmetric phase. The discontinuity in $\langle\Tr M^\dagger M\rangle$ along the critical line diminishes, as $d_3$ gets larger (see Fig.\ref{latent}), but it seems that the discontinuity persists, even if its magnitude diminishes in the limit $d_3\rightarrow\infty$ suggesting that there is a first order phase transition for any (positive) value of $d_3$.

\begin{table}
\begin{center}
\begin{tabular}{lll}
\hline
$\beta$ & Lattice volumes \\ 
\hline
12 & $8^3$, $10^3$, $12^3$, $16^3$  \\ 
16 & $12^3$, $16^3$, $20^3$\\
20 & $16^3$, $20^3$, $24^3$\\
\hline
\end{tabular}
\caption{Lattices used in the continuum and thermodynamical extrapolation of the critical point, seen in Fig.\ref{finite_size}.}
\label{lattices}
\end{center}
\end{table}

\section{Conclusions}
In this paper, exact relations between the lattice and continuum $\MSb$ regulated formulations of the Z(3)-symmetric super-renormalizable effective theory of hot QCD, defined by Eqs.(\ref{action}),(\ref{V0}), and (\ref{V1}), have been calculated. The Lagrangians and the operators up to cubic ones have been matched to $\mathcal{O}(a^0)$. These results make the non-perturbative lattice study of the theory possible. 

An interesting model  with non-trivial dynamics is obtained by setting $c_i=0$ in Eq.(\ref{V0}). The model amounts to a natural generalization of EQCD to complex variables.
The phase diagram of the model has been determined using lattice simulations. Two distinct phases were found, a symmetric phase with small $\langle\Tr M^\dagger M\rangle_{\MSb}$, and a broken phase with large $\langle\Tr M^\dagger M\rangle_{\MSb}$. The two phases were found to be separated by a strong first order transition with a large surface tension and discontinuities in the operators. In contrary to EQCD, where the first order line terminates at a tricritical point, the model seems to have a first order transition with all values of $d_2$ and $d_3$.

In the future, it is our goal to map out the phase 
diagram in the full parameter space of the model, rather
than in a restricted region as in the present exploratory 
study, in order to search for regions in which
the phase diagram would resemble that expected for 
the finite-temperature SU(3) pure Yang-Mills theory. 

\section*{Acknowledgments}
The author thanks K.~Kajantie for suggesting
this topic and for numerous comments concerning the text.
The author also thanks M.~Laine, K.~Rummukainen, Y.~Schr\"oder and A.~Vuorinen for invaluable advice. 
 This research has been supported by Academy of
Finland, contract number 109720 and the EU I3 Activity
RII3-CT-2004-506078 HadronPhysics. Simulations were carried out at CSC - Scientific Computing Ltd., Finland; the total amount of computing
power used was $\sim1\times10^{16}$ flops.

\appendix
\section{Details of renormalization}
In this appendix, we give details of the calculation of the renormalization of the mass parameters $\hat{c}_1$ and $\hat{d}_1$ and the condensates $\langle g_3^{-2}\Tr Z^\dagger Z\rangle$, $\langle g_3^{-2}\Tr M^\dagger M\rangle$, $\langle g_3^{-2}\Tr Z \Tr Z^\dagger \rangle$,$\langle g_3^{-3}2\Det Z\rangle$, and $\langle g_3^{-3}2\myRe\Tr M^3 \rangle$. 

The renormalization calculation compares ultraviolet properties of the two regularizations and thus it is irrelevant, in which phase we carry out the computation. We chose to work around the symmetric vacuum, since the Feynman rules are the simplest this way. However, in this vacuum, all components of the gluon are massless and one therefore has to deal with infrared divergences.
The infrared divergences in the two regularizations are the same, and cancel exactly in the final results. 

 Using the expansion (\ref{expansion}), the potential $V_2$ is a function of $H$ and $A$ only. The unit matrix commutes with any SU(3) matrix, and the interaction with the gluon field arises from commutator in the covariant derivative in adjoint representation. Thus the gluons couple, on tree-level, only to $H$ and $A$ 

Using this expansion, there are four tree-level mass terms in the Lagrangian that require renormalization:
\begin{equation}
	\begin{array}{ll}
 \frac{1}{2} \hat{c}_1 \phi^2,&  \hlf(\hat{c}_1+\hat{d}_1)  H^aH^a\\
 \frac{1}{2}\hat{c}_1 \chi^2, & \hlf(\hat{c}_1+\hat{d}_1)  A^aA^a.
\end{array}
\end{equation}
The coefficients $\hat{c}_1$ and $\hat{d}_1$ have to be adjusted such that both regularization schemes give the same physical masses for all fields $\phi$, $\chi$, $H$ and $A$. The theory is super-renormalizible, and there are divergent contributions up to two-loop level only. The masses are obtained from the low momentum properties of the two-point correlator:
\begin{align}
\lim_{k,p\rightarrow 0}\langle\langle \phi(k) \phi(p) \rangle \rangle&= \delta^{(3)}(k+p) \frac{1}{k ^2 + m^2}
\end{align}
The difference between the correlators in the two schemes is in the mass and the wave function renormalization. However, the effect of the wave function renormalization is of order $\mathcal{O}(a)$ and can be neglected. 
To get the same masses in the different schemes, we enforce the condition that the two-point correlators give identical values in the low momentum limit. The zero-momentum lattice correlator can be written in a weak-coupling expansion
\begin{align}
\langle\langle \phi(0) \phi(0) \rangle\rangle_{a}
=& \left[ \langle \phi(0)\phi(0) \rangle_a - \langle \phi(0)\phi(0)\rangle_{\MSb} \right] + \left[\langle \phi(0) \phi(0) (-S_I) \rangle_a - \langle \phi(0) \phi(0) (-S_I) \rangle_{\MSb}\right] \nonumber \\&+ \frac{1}{2}\left[ \langle \phi(0) \phi(0) S_I^2 \rangle_a - \langle \phi(0) \phi(0) S_I^2 \rangle_{\MSb} \right] + \langle \langle\phi(0) \phi(0) \rangle\rangle_{\MSb} + \mathcal{O}(a) ,
\end{align}
where double and single brackets represent exact and Gaussian expectation values, respectively, and subscripts give the regularization scheme. From here, we can read the condition for the lattice mass term $\hat{c}_1$ by requiring that the exact correlators give the same value up to order $\mathcal{O}(\beta^{-1})$:
\begin{align}
\hat{c}_1 = \frac{c_1}{g_3^4}+\frac{1}{g_3^4} \bigg( \left[\langle \phi \phi (-S_I) \rangle_{a,1\textrm{PI}} - \langle \phi \phi (-S_I) \rangle_{\MSb ,1\textrm{PI}}\right] \nonumber\\+ \frac{1}{2}\left[ \langle \phi \phi S_I^2 \rangle_{a,1\textrm{PI}} - \langle \phi \phi S_I^2 \rangle_{\MSb,1\textrm{PI}} \right] \bigg)  + \mathcal{O}(a)
\end{align}
Similarly, we get for the mass term of adjoint fields:
\begin{align}
\hat{d}_1 = \left[\frac{c_1}{g_3^4}-\hat{c}_1 \right] + \frac{d_1}{g_3^4} + \frac{1}{g_3^4} \bigg( \left[\langle HH (-S_I) \rangle_{a,1\textrm{PI}} - \langle HH (-S_I) \rangle_{\MSb ,1\textrm{PI}}\right] \nonumber \\+ \frac{1}{2}\left[ \langle HH S_I^2 \rangle_{a,1\textrm{PI}} - \langle HH S_I^2 \rangle_{\MSb,1\textrm{PI}} \right] \bigg)  + \mathcal{O}(a)
\end{align}
The correlators in both regularizations are in two-loop weak-coupling expansion infrared-divergent quantities. However,  since the infrared properties of the two regularization schemes are the same, the infrared divergences cancel exactly in the difference.

The renormalization of the condensates is done very similarly. The condensates can be expressed, in both regularization schemes, as derivatives with respect to mass parameters of the free energy and thus they can be related. For the quadratic condensates we get:
\begin{align}
 \langle \Tr Z^\dagger Z \rangle_{\MSb} &= \frac{\partial f_{\MSb}}{\partial c_1(\mu)}=\langle \Tr Z^\dagger Z \rangle_a + \frac{\partial (f_{\MSb}-f_a)}{\partial c_1(\mu)} \\
\langle  \Tr M^\dagger M \rangle_{\MSb} &= \frac{\partial f_{\MSb}}{\partial d_1(\mu)}=\langle \Tr M^\dagger M \rangle_a +\frac{\partial (f_{\MSb}-f_a)}{\partial d_1(\mu)},
\end{align}
and for the cubic:
\begin{align}
 \langle 2\myRe\Det Z\rangle_{\MSb} &=\frac{\partial f_{\MSb}}{\partial c_2}= \langle 2\myRe\Det Z\rangle_{a}+ \frac{\partial (f_{\MSb}-f_a)}{\partial c_2}\\
 \langle 2\myRe\Tr M^3 \rangle_{\MSb} &=\frac{\partial f_{\MSb}}{\partial d_2}= \langle 2\myRe\Tr M^3 \rangle_{a}+ \frac{\partial (f_{\MSb}-f_a)}{\partial d_2}.
\end{align}
Due to the super-renormalizability the difference in free energy is dimensionally of the form:
\begin{align}
f_{\MSb}-f_a&=\frac{1}{4\pi}\left[ A_{1,0}\frac{1}{a^3}+D_{1,0}\frac{c_1(\bar{\mu})}{a}+D_{1,1}\frac{d_1(\bar{\mu})}{a}\right]\nonumber\\
&+ \frac{1}{(4\pi)^2}\Big[B_{2,0} \frac{g_3^2}{a^2}+C_{2,1}\frac{c_3}{a^2} +C_{2,2}\frac{d_3}{a^2}\nonumber\\
&\qquad\qquad + E_{2,0} c_2^2 + E_{2,1} d_2^2 + E_{2,2} c_2d_2 \nonumber \\
&\qquad \qquad + D_{2,0}g_3^2c_1(\bar{\mu})+ D_{2,1}c_3c_1(\bar{\mu})+ D_{2,2}d_3c_1(\bar{\mu}) \nonumber\\ 
&\qquad\qquad  + D_{2,3}g_3^2d_1(\bar{\mu})+ D_{2,4}c_3d_1(\bar{\mu})+ D_{2,5}d_3d_1(\bar{\mu}) \Big] \nonumber\\
&+\frac{1}{(4\pi)^3}\Big[B_{3,0} \frac{g_3^4}{a}+C_{3,1}\frac{g_3^2 c_3}{a}+C_{3,2}\frac{g_3^2 d_3}{a} + C_{3,3}\frac{c_3^2}{a} + C_{3,4}\frac{d_3^2}{a}+C_{3,5} \frac{c_3d_3}{a}\Big] \nonumber \\
&+\frac{1}{(4\pi)^4}\Big[B_{4,0}g_3^6+C_{4,1}g_3^4 c_3 + C_{4,2}g_3^4 d_3 +C_{4,3}g_3^2 c_3^2 + C_{4,3}g_3^2 d_3^2 + C_{4,4}g_3^2 c_3 d_3 \nonumber \\
&\qquad \qquad + C_{4,5}c_3^3 + C_{4,6}d_3^3 + C_{4,6}c_3^2 d_3 + C_{4,7}c_3 d_3^2\Big]+\mathcal{O}(a),\label{dimanalysis}
\end{align}
where the dimensionless coefficients $A_{i,j}$, $B_{i,j}$, $C_{i,j}$, $D_{i,j}$, and $E_{i,j}$ are functions of a dimensionless combination  $a\bar{\mu}$ only. The coefficients $C_{i,j}$ and $D_{i,j}$ follow from an $i$-loop computation. For the quadratic and cubic condensates we need to know coefficients $D_{i,j}$ and $E_{i,j}$ in order to obtain the matching of the condensates to order $\mathcal{O}(a^0)$, which follow from a two-loop calculation:
\begin{align}
  D_{1,0}&=-\Sigma\Nc^2\nonumber\\
  D_{1,1}&=-\Sigma(\Nc^2-1)\nonumber\\
  D_{2,0}&=-2\Nc(\Nc^2-1)\left( \ln\frac{6}{a\bar{\mu}}+\zeta+\frac{\Sigma}{4}-\delta\right)\nonumber\\
  D_{2,1}&=0\nonumber\\
D_{2,2}&=0\nonumber\\
  D_{2,3}&=-2\Nc(\Nc^2-1)\left( \ln\frac{6}{a\bar{\mu}}+\zeta+\frac{\Sigma}{4}-\delta\right)\nonumber\\
  D_{2,4}&=0\nonumber\\
D_{2,5}&=0\nonumber\\
  E_{2,0}&=-\left[\left(\frac{2}{9}+\frac{1}{6}(\Nc^2-1)\right)+\frac{4}{3}\left(\frac{1}{\Nc}-\frac{5}{4}\Nc+\frac{1}{4}\Nc^3\right)\right]\left( \ln\frac{6}{a\bar{\mu}}+\zeta\right)\nonumber\\
  E_{2,1}&=-12\left(\frac{1}{\Nc}-\frac{5}{4}\Nc+\frac{1}{4}\Nc^3\right)\left( \ln\frac{6}{a\bar{\mu}}+\zeta\right)\nonumber\\
  E_{2,2}&=-8\left(\frac{1}{\Nc}-\frac{5}{4}\Nc+\frac{1}{4}\Nc^3\right)\left( \ln\frac{6}{a\bar{\mu}}+\zeta\right).
\end{align}
For the quartic condensates, however, the coefficients $C_{i,j}$ are needed and a four-loop lattice perturbation theory calculation is required for the matching. For the gluon condensates, also the $B_{i,j}$ are needed. The coefficients $B_{2,0}$ and $B_{3,0}$ have been calculated in \cite{Heller:1984hx} and \cite{Panagopoulos:2006ky}, respectively. The coefficient $B_{4,0}$ has been calculated for $\Nc=3$ using stochastic perturbation theory in \cite{DiRenzo:2006nh}.

\section{Feynman rules}
Using the expanded fields, the potentials become:
\begin{align}
V_0(Z) = &g_3^4 \left\{\frac{\hat{c}_1}{2} \phi^2 
+ \frac{\hat{c}_1}{2} \chi^2 
+ \hat{c}_1  \Tr[A \cdot A]  
+ \hat{c}_1  \Tr[H \cdot H]\right\} 
\nonumber\\
+& g_3^3 \bigg\{\frac{1}{3!} \frac{2 \hat{c}_2}{\sqrt{6}} \phi^3 
- \frac{1}{2!} \frac{2 \hat{c}_2}{\sqrt{6}} \phi \chi^2 
+ \frac{1}{2!} \frac{2 \hat{c}_2}{\sqrt{6}} \phi \Tr[A \cdot A] 
- \frac{1}{2!}\frac{2 \hat{c}_2}{\sqrt{6}} \phi  \Tr[H \cdot H] \nonumber \\ 
+ &  \frac{1}{3!} 4 \hat{c}_2 \Tr[ H \cdot H \cdot H] 
+ \frac{2 \hat{c}_2}{\sqrt{6}} \chi \Tr[A \cdot H] 
- \frac{1}{2!} 4 \hat{c}_2\Tr[A \cdot A \cdot H] \bigg\}
\nonumber \\
+  &g_3^2\bigg\{ \frac{1}{4!} 2 \hat{c}_3 \phi^4 
+ \frac{1}{4!} 2 \hat{c}_3 \chi^4   
+  \frac{1}{2!2!} \frac{2}{3} \hat{c}_3 \phi^2 \chi^2  \nonumber \\
+ & \frac{1}{2!2!} 4 \hat{c}_3 \phi^2 \Tr[H \cdot H] 
+ \frac{1}{2!2!} \frac{4}{3} \hat{c}_3 \chi^2 \Tr[H \cdot H]  
+ \frac{1}{2!2!}  \frac{4}{3} \hat{c}_3 \phi^2 \Tr[A \cdot A] \nonumber \\
+ & \frac{1}{2!2!} 4 \hat{c}_3\chi^2 \Tr[A \cdot A] 
+  \frac{4}{3}\hat{c}_3 \phi \chi \Tr[A \cdot H]  \nonumber \\
+ & \frac{1}{3!} 4\sqrt{6}  \hat{c}_3 \phi \Tr[H \cdot H \cdot H] 
+ \frac{1}{3!} 4\sqrt{6}  \hat{c}_3 \chi \Tr[A \cdot A \cdot A]\nonumber \\
+& \frac{1}{2!}\frac{4}{3}\sqrt{6}  \hat{c}_3 \phi \Tr[A \cdot A \cdot H] 
+ \frac{1}{2!}\frac{4}{3}\sqrt{6} \hat{c}_3 \chi \Tr[A \cdot H \cdot H] \nonumber \\
+ & \frac{1}{4!} 24 \hat{c}_3 \Tr[A \cdot A \cdot A \cdot A] 
+ \frac{1}{4!} 24  \hat{c}_3 \Tr[H \cdot H \cdot H \cdot H]\nonumber \\
+& \frac{1}{2!2!} 16  \hat{c}_3 \Tr[A \cdot A \cdot H \cdot H]  
- \frac{1}{2!2!} 8\hat{c}_3 \Tr[A \cdot H \cdot A \cdot H]\bigg\}
\end{align}
and 
\begin{align}
V_1(Z) =  & g_3^4\left\{ \hat{d}_1 \Tr[A \cdot A] + \hat{d}_1 \Tr[H \cdot H]\right\} \nonumber \\ 
 +&   g_3^3\left\{-6 \hat{d}_2 \Tr[A \cdot A \cdot H] + 2 \hat{d}_2 \Tr[H \cdot H \cdot H] \right\}\nonumber \\
+ &  g_3^2 \Big\{ \hat{d}_3 \Tr[A \cdot A \cdot A \cdot A] + 4 \hat{d}_3 \Tr[A \cdot A \cdot H \cdot H] 
\nonumber \\ - &   2 \hat{d}_3\Tr[A \cdot H \cdot A \cdot H] + \hat{d}_3 \Tr[H \cdot H \cdot H \cdot H] \Big\}.
\end{align}

The gauge part of the scalar Lagrangian in Fourier space (momentum conservation, all integrations over Brillouin zone, with measure $\latint{p}$, and sums understood) becomes \cite{Laine:1995ag}:
\begin{align}
S_Z &= i\frac{g_3}{2!} \, f^{abc} \, (\widetilde{p-q})_i A^a(p)A^b(q)A^c_i(r) 
+ \frac{g_3^2}{2!2!}  2f^{ace} f^{bde}(\wideundertilde{p-q})_i \delta_{ij} A^a(p)A^b(q)A^c_i(r)A_j^d(s)\nonumber  \\
    &+ i\frac{g_3}{2!} \, f^{abc} \, (\widetilde{p-q})_i H^a(p)H^b(q)A^c_i(r) 
+ \frac{g_3^2}{2!2!}  2f^{ace} f^{bde}  (\wideundertilde{p-q})_i \delta_{ij} H^a(p)H^b(q)A^c_i(r)A_j^d(s), 
\end{align}
where we use a compact notation:
\begin{align}
\undertilde{p_i}=\cos\frac{a p_i}{2}, \quad\quad  \widetilde{p_i}=\frac{2}{a}\sin\frac{a p_i}{2},\quad \textrm{and} \quad \widetilde{p}^2=\frac{4}{a^2}\sum_i\sin^2\frac{ap_i}{2} 
\end{align}
In addition to these there is the pure gluon and gauge fixing sector \cite{Laine:1997dy}
\begin{align}
S_W &= \frac{1}{2}\widetilde{p}^2A^a_i(-p)A^a_i(p)+\widetilde{p}^2\bar{c}^a(p)c^a(p)+i g_3 f^{abc} \undertilde{r_i}\tilde{p}_i \bar{c}^a(p) c^c(r)A_i^b(q) \nonumber \\
-& \frac{1}{24}g_3^2 a^2 (f^{ace}f^{bde}+f^{ade}f^{bce}) \widetilde{s}_i \widetilde{p}_i \bar{c}^a(p)A_i^c(q)A_i^d(r)c^b(s)+ \frac{1}{2}g_3^2\frac{\Nc}{12 a}A^a_i(-p)A^a_i(p)+ S_3 + S_4.
\end{align}
Contributions of three and four gluon vertices $S_3$ and $S_4$ can be found in \cite{Rothe:2005nw}, Eqs. (15.39),(15.43) and (15.53), where one needs to replace $(\frac{2}{3})(\delta_{AB}\delta_{CD}+\ldots)$ with $(\frac{2}{\Nc})(\delta_{AB}\delta_{CD}+\ldots)$ \cite{Laine:1997dy}.

\section{Calculation of the diagrams}
The perturbation theory calculations were done using symbolic manipulation language FORM\cite{Vermaseren:2000nd}.
For formalized computation, it is advantageous to write
all the color tensors in the fundamental representation, i.e. using the generators of the group:
\begin{align}
\Tr[T^a T^b T^c] &= \frac{1}{4} ( d^{abc} + i f^{abc} ) \\
\Tr[T^a T^b]     &= \frac{1}{2} \delta^{ab}.
\end{align}
Then all the color contractions in loop calculations can be done systematically with repeated use of the Fiertz identity:
\be
T^{a}_{ij}T^{a}_{kl} = \hlf ( \delta_{il} \delta_{jk} - \frac{1}{\Nc} \delta_{ij} \delta_{kl} ).
\ee
The following combinations are found in the action:
\begin{align}
i f^{abc} &= 2 \Tr(T^a [T^b,T^c]) \\
d^{abc}   &= 2 \Tr(T^a \{T^b,T^c\} )\\
f^{abc} f^{bde} &= -2 \Tr[T^a,T^c] \Tr[T^b,T^d] \\
d^{abe} d^{cde} &= 2 \Tr\{T^a,T^b\} \Tr\{T^c,T^d\} - \frac{2}{N_c}\delta^{ab} \delta^{cd}
\end{align}
In lattice perturbation theory, the numerators of the integrals contain complex trigonometric objects. These can be systematically reduced to squares of sines, which also appear in the denominator, by repeated use of the following formulae (no summation over repeated indices):
\begin{align}
 \widetilde{x + y}_i &= \widetilde{x}_i\undertilde{y_i} + \widetilde{y_i}\undertilde{x_i}\\
 \wideundertilde{(x+y)_i} &= \undertilde{x_i}\undertilde{y_i}-\frac{a^2}{4} \widetilde{x}_i\widetilde{y}_i \\
\undertilde{x_i}^2 &= \delta_{ii}-\frac{a^4}{4}\widetilde{x}_i^2 \\
\widetilde{x}_i\widetilde{y}_i\undertilde{x_i}\undertilde{y_i} &= \frac{1}{2} \left((\widetilde{x+y})_i^2-\widetilde{x}_i^2\undertilde{x_i}\undertilde{y_i}
-\widetilde{y}_i^2\undertilde{x_i}\undertilde{y_i}\right).\label{trigs}
\end{align}
This procedure generalizes trivially also to higher order loop calculations.
The set of integrals can be further reduced by applying a trigonometric identity (for $j\geq2$):
\be
\frac{a^2}{4}\latint{x}\frac{\sum_i \widetilde{x}_i^4 }{(\widetilde{x}^2+m^2)^j}=\frac{1}{j-1}\latint{x}\frac{\frac{a^2}{4}\widetilde{x}^2-\frac{3}{2}}{(\widetilde{x}^2+m^2)^{j-1}}+\latint{x}\frac{\widetilde{x}^2}{(\widetilde{x}^2+m^2)^j}.\label{alpha}
\ee

\section{Diagrams for mass renormalization}
In this section, we give the zero momentum diagrams that affect the mass renormalization. The expressions are in lattice regularization and the symbol ''$\MSb$'' refers to the result of the corresponding diagram in the $\MSb$ regularization. Solid and wiggly lines represent scalars and gluons, respectively. Symbols in parentheses indicate fields running in the internal scalar lines. The symmetry factors are included in the coefficients.

 The following diagrams with zero incoming momenta contribute to the renormalization of the mass term $\hat{c}_1$ of $\phi$-field (with external lines $\phi$):
\begin{itemize}
\item One-loop: The mass $m_i$ refers to the mass of the field running in the loop. In the difference between continuum and lattice regularization, the mass dependence cancels.

\begin{align}
\begin{array}{c}
\vspace{-8mm}
\includegraphics[width = 80pt,height = 60pt]{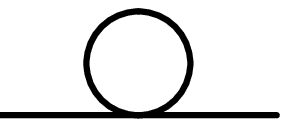}
\end{array}
  &= 
\left\{ \begin{array}{ccl}
(\phi) & : & -1\\
(\chi) & : & -\frac{1}{3}\\
(A)    & : & -\frac{1}{3}(\Nc^2-1) \\
(H)    & : & -(\Nc^2-1) \\
\end{array} \right\} g_3^2 \hat{c}_3 I(m_i)\nonumber\\&=-\frac{4}{3}\Nc^2 g_3^2 \hat{c}_3 \frac{\Sigma}{4 \pi a} + \cont +\MSb
\end{align}
\item{Two-loop:}
\begin{itemize}
\item Terms proportional to $\hat{c}_3^2 g_3^4$:  Masses $m_1$, $m_2$ and $m_3$ in the denominator refer to the masses of the internal lines and $m_d^2 = g_3^4(\hat{c}_1+\hat{d_1})$. 
\begin{align}
\begin{array}{c}
\vspace{-2mm}
\includegraphics[width = 80pt,height = 60pt]{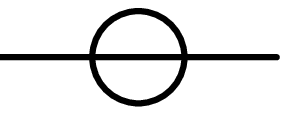}
\end{array}
 &  = \begin{array}{cl} 
\left\{ \begin{array}{ccl} 
(\phi \phi \phi )& : &\frac{2}{3} \\
(\chi \chi \phi) & : &\frac{2}{9}  \\
(A A \phi  )     & : &\frac{2}{9}(\Nc^2 -1)\\
(H H \phi )      & : &2(\Nc^2 -1) \\
(\chi A H)       & : & \frac{4}{9}(\Nc^2 -1)\\
(A A H )         & : & -\frac{1}{3}\Nc(\Nc^2 -1)\\
(H H H)          & : & -\Nc(\Nc^2 -1)
\end{array} \right\} g_3^4 \hat{c}_3^2
\end{array}\nonumber
\\ &\times H(m_1,m_2,m_3)\nonumber 
\end{align}
\be
 \hspace{-5mm} =\left( -\frac{16}{9} +\frac{4}{3}\Nc + \frac{24}{9}\Nc^2 -\frac{4}{3}\Nc^3 \right) \frac{g_3^4 \hat{c}_3 ^2}{16 \pi^2} \left[ \log\frac{6}{a \bar{\mu}} + \zeta\right] + \cont +\MSb
\ee
\item Terms proportional to $\hat{c}_3 g_3^4$: The tadpoles are cancelled by the 1-loop counter terms.
\begin{align}
\begin{array}{c}
\vspace{-10mm}
\includegraphics[width = 80pt,height = 60pt]{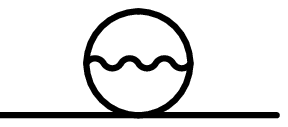}
\end{array}
&
=\left\{ \begin{array}{ccl} 
(A) & : &  -\frac{1}{3} \Nc (\Nc^2-1)\\
(H) & : &  -\Nc (\Nc^2-1)
\end{array}\right\}g_3^4 \hat{c}_3 \nonumber
\end{align}
\begin{align}
&\times\left[2H(m_d,m_d,0)+\left(2I(0)-I(m_d)\right)\left(-\partial_{m_d^2}I(m_d)\right)+4m_d^2 H'(m_d,m_d,0)-a^2 G(m_d,m_d)\right] \nonumber\\
&=-\left(\frac{8}{3}\Nc(\Nc^2-1)\right)\frac{g_3^4\hat{c}_3}{16\pi^2}\left[ \log\frac{6}{a \bar{\mu}} + \zeta - \delta\right]  + \cont+ \tadpoles +\MSb\end{align}
\begin{align}
  \begin{array}{c}
\vspace{-10mm}
\includegraphics[width = 60pt,height = 80pt]{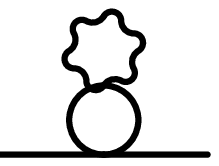}
\end{array} &= \left\{ \begin{array}{ccl}
 (A) & : & \frac{1}{3}\Nc (\Nc^2 -1) \\
 (H) & : & \Nc (\Nc^2-1) \end{array}\right\}\hat{c}_3g_3^4 \nonumber
\end{align}
\begin{align}
&\times\left[ 3I(0)(-\partial_{m_d^2})I(m_d)+a^2\left\{\frac{m_d^2}{2} I(0)\left(-\partial_{m_d^2}I(m_d)\right)-\frac{1}{2}I(0)I(m_d)\right\}\right]
 \nonumber
\\&= -\left( \frac{8}{3}\Nc(\Nc^2-1) \right)  \frac{g_3^4 \hat{c}_3}{16 \pi^2} \frac{\Sigma^2}{4} + \cont + \tadpoles +\MSb
\end{align}
\end{itemize}
\end{itemize}

The following diagrams with zero incoming momenta contribute to the renormalization of the mass term $\hat{d}_1$ of $H$-field (with external lines $H$):
\begin{itemize}
 \item One-loop diagrams: 
\begin{align}
\begin{array}{c}
\vspace{-8mm}
\includegraphics[width = 80pt,height = 60pt]{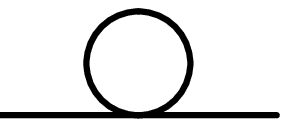}
\end{array}
  &= 
\left\{ \begin{array}{ccl}
(\phi) & : & -\hat{c}_3\\
(\chi) & : & -\frac{1}{3}\hat{c}_3\\
(A)    & : & -(2\Nc-1/\Nc)(\hat{c}_3+\hat{d}_3) \\
(H)    & : & -(2\Nc-3/\Nc)(\hat{c}_3+\hat{d}_3) \\
\end{array} \right\} g_3^2
I(m_i)\nonumber\\
&=\left[ -\frac{4}{3}\hat{c}_3 - 4(\Nc-1/\Nc)(\hat{c}_3+\hat{d}_3) \right]g_3^2  \frac{\Sigma}{4 \pi a} + \cont +\MSb
\\
\begin{array}{c}
\vspace{-8mm}
\includegraphics[width = 80pt,height = 60pt]{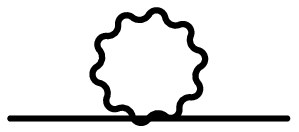}
\end{array}
  &= -3g_3^2 \Nc I(0) =-3g_3^2 \Nc \frac{\Sigma}{4 \pi a } + \cont+\MSb
\\
\begin{array}{c}
\vspace{-8mm}
\includegraphics[width = 80pt,height = 60pt]{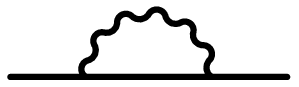}
\end{array}
  &= g_3^2 \Nc  I(0) = g_3^2 \Nc \frac{\Sigma}{4 \pi a }+\cont+\MSb
\end{align}	
\item Two-loop diagrams:
\begin{itemize} \item Terms proportional to $g_3^4\hat{c}_3^2$:
\begin{align}
\begin{array}{c}
\vspace{-2mm}
\includegraphics[width = 80pt,height = 60pt]{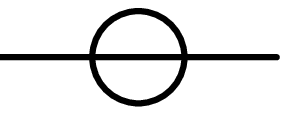}
\end{array}
 &  =  
\left\{ \begin{array}{ccl} 
(\phi \phi H )& : & 2 \hat{c}_3^2\\
(\phi \chi A) & : & \frac{4}{9} \hat{c}_3^2 \\
(A A \phi  )     & : &-\frac{1}{3} \Nc \hat{c}_3^2 \\
(H H \phi )      & : &-3\Nc\hat{c}_3^2 \\
(\chi \chi H)       & : & \frac{2}{9} \hat{c}_3^2 \\
(\chi A H )         & : & -\frac{2}{3}\Nc \hat{c}_3^2\\
(A A H)          & : & -2(1-3/\Nc^2 - \frac{3}{2}\Nc^2)(\hat{c}_3+\hat{d}_3)^2 \\
(H H H)          & : & -6(1-3/\Nc^2 - \frac{1}{6}\Nc^2)(\hat{c}_3+\hat{d}_3)^2 \\
\end{array} \right\} g_3^4 H(m_1,m_2,m_3)\nonumber
\end{align}
\begin{align}
&= \left[ (\frac{8}{3}-4 N_c)\hat{c}_3^2 +(4\Nc^2+24/\Nc^2-8)(\hat{c}_3+\hat{d}_3)^2 \right]\times \frac{g_3^4}{16 \pi^2} \left[ \log\frac{6}{a \bar{\mu}} + \zeta\right]+\cont+\MSb
\end{align}
\item Terms proportional to $g_3^4(\hat{c}_3+\hat{d}_3)$:
\begin{align}
\begin{array}{c}
\vspace{-10mm}
\includegraphics[width = 80pt,height = 60pt]{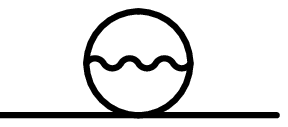}
\end{array}
&
=\left\{ \begin{array}{ccl} 
(A) & : &  -2(\Nc^2-\frac{1}{2}) \\
(H) & : &  -2(\Nc^2-\frac{3}{2})
\end{array}\right\}g_3^4 (\hat{c}_3 +\hat{d}_3 )\nonumber
\end{align}
\begin{align}
 &\times\left[2H(m_d,m_d,0)+\left(2I(0)-I(m_d)\right)\left(-\partial_{m_d^2}I(m_d)\right)+4m_d^2 H'(m_d,m_d,0)-a^2 G(m_d,m_d)\right] \nonumber\\
&=-8(\Nc^2-1) (\hat{c}_3+\hat{d}_3)\frac{g_3^4}{16 \pi^2} \left[ \log\frac{6}{a \bar{\mu}} + \zeta - \delta \right] + \cont + \tadpoles+\MSb
\end{align}
\begin{align}
  \begin{array}{c}
\vspace{-10mm}
\includegraphics[width = 60pt,height = 80pt]{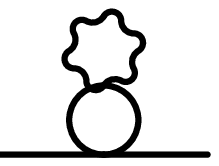}
\end{array} &= \left\{ \begin{array}{ccl}
 (A) & : & -2(\Nc^2-\frac{1}{2}) \\
 (H) & : & -2(\Nc^2-\frac{3}{2}) \end{array}\right\} g_3^4 (\hat{c}_3+\hat{d}_3^2)\nonumber
\end{align}
\begin{align}
\times &\left[3I(0)(-\partial_{m_d^2})I(m_d)+a^2\left\{\frac{m_d^2}{2} I(0)\left(-\partial_{m_d^2}I(m_d)\right)-\frac{1}{2}I(0)I(m_d)\right\}\right]\nonumber\\
&= -2(\Nc^2-1) g_3^4 (\hat{c}_3+\hat{d}_3)
\frac{\Sigma^2}{16\pi^2} +\cont+\MSb
\end{align}

\item Terms proportional to $g_3^4$: The coupling of the adjoint fields $A$ and $H$ is exactly the same as as in EQCD, so 
the term proportional to $g_3^4$ can be taken from EQCD \cite{Laine:1997dy}. However, at two-loop level diagrams with an adjoint scalar loop contribute two times
since there are two adjoint fields. The diagrams with adjoint loops are the following:
\begin{align}
  \begin{array}{c}
\vspace{-10mm}
\includegraphics[width = 60pt,height = 80pt]{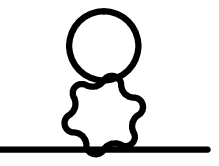}
\end{array} &= \Nc^2 g_3^4 \Latint{x}{y} \frac{\undertilde{2y_i}}{\widetilde{x}^2\widetilde{x}^2(\widetilde{y}^2+m_d^2) } \nonumber
\end{align}
\begin{align}
=& \Nc^2 g_3^4 \Big[ 3 \Latint{x}{y} \frac{1}{\widetilde{x}^2 \widetilde{x}^2(\widetilde{y}^2+m_d^2)} \nonumber \\
  &+ \frac{a^2}{2} \Latint{x}{y} \frac{1}{\widetilde{x}^2 \widetilde{x}^2} \nonumber \\
  &+  \frac{a^2 m_d^2}{2}\Latint{x}{y} \frac{1}{\widetilde{x}^2 \widetilde{x}^2(\widetilde{y}^2+m_d^2)}\Big] 
\end{align}
\begin{align}
\begin{array}{c}
\vspace{-10mm}
\includegraphics[width = 80pt,height = 60pt]{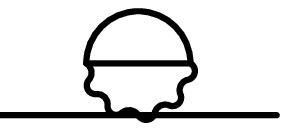}
\end{array}
&= -\frac{1}{2}\Nc^2 g_3^4  \frac{\widetilde{(2x+y)}^2}{(\widetilde{x}^2 + m_d^2) \widetilde{y}^2\widetilde{y}^2 (\widetilde{(x+y)}^2 +m_d^2)} \nonumber
\end{align}
\begin{align}
= \Nc^2 g_3^4 \Big[  & + \frac{1}{2}H(m_d,m_d,0) \nonumber \\
		     &- 2 \Latint{x}{y} \frac{1}{\widetilde{x}^2 \widetilde{x}^2} \prop{y}{m_d}  \nonumber\\
		     &+2 m_d^2 \Latint{x}{y}\frac{1}{\widetilde{x}^2 \widetilde{x}^2} \prop{y}{m_d}\prop{(x+y)}{m_d}  \nonumber \\
		     &-\frac{a^2}{2} \Latint{x}{y} \frac{\sum_i \widetilde{x}_i^2 \widetilde{y}_i^2}{[ \widetilde{(x+y)}\!\,^2 ] ^2 (\widetilde{x}^2 + m_d^2)(\widetilde{y}^2 + m_d^2)}\Big]
\end{align}
The last line can be written in a more familiar form using the definition of $\rho$, Eq.(\ref{rho}), and the trigonometric identity Eq.(\ref{alpha}):
\begin{align}
\Latint{x}{y} \frac{\sum_i \widetilde{x}_i^2 \widetilde{y}_i^2}{[ \widetilde{(x+y)}\!\,^2 ] ^2 (\widetilde{x}^2 + m_d^2)(\widetilde{y}^2 + m_d^2)}&   \nonumber\\
= \frac{\rho}{4 \pi^2 a^2} + \Latint{x}{y}\frac{1}{\widetilde{x}^2\widetilde{x}^2} -& \frac{2}{a^2} \Latint{x}{y}\frac{1}{\widetilde{x}^2}\frac{1}{\widetilde{y}^2\widetilde{y}^2} + \cont.
\end{align}
The infrared divergences in these two diagrams cancel and the sum of the diagrams becomes:
\begin{align}
\begin{array}{c}
\vspace{-10mm}
\includegraphics[width = 60pt,height = 80pt]{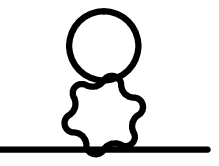}
\end{array}
+
\begin{array}{c}
\vspace{-10mm}
\includegraphics[width = 80pt,height = 60pt]{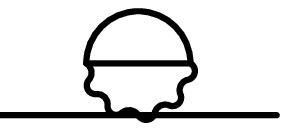}
\end{array}
 = \Nc^2 g_3^4 \frac{1}{16\pi^2}\left(\frac{1}{2} \left[ \log\frac{6}{a \bar{\mu}} + \zeta \right] + 2\rho \right) + \cont+\MSb
 \end{align}
There are also two other diagrams:
\begin{align}
\begin{array}{c}
\vspace{-10mm}
\includegraphics[width = 60pt,height = 80pt]{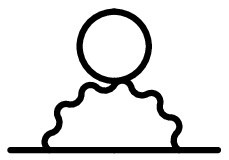}
\end{array} &= - \Nc^2 g_3^4 \Latint{x}{y} \frac{\sum_i\widetilde{x}^2_i \undertilde{(2y)}^2 \!\,_i}{\widetilde{x}^2\widetilde{x}^2(\widetilde{x}^2+m_d^2)(\widetilde{y}^2+m_d^2)}
\\
\begin{array}{c}
\vspace{-10mm}
\includegraphics[width = 80pt,height = 70pt]{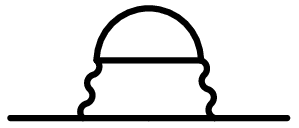}
\end{array} &= \Nc^2 g_3^4 \frac{1}{2}\Latint{x}{y}\frac{\sum_i \widetilde{x}_i\widetilde{(x+2y)}_i \sum_j \widetilde{x}_j\widetilde{(x+2y)}_j}{\widetilde{x}^2\widetilde{x}^2(\widetilde{x}^2+m_d^2)(\widetilde{y}^2+m_d^2)(\widetilde{(x+y)}^2+m_d^2)}.
 \end{align}
After repeated use of Eqs.(\ref{trigs}), the both diagrams can be written in the form
\be
\pm \Nc^2 g_3^4 \Latint{x}{y} \frac{ \widetilde{x}^2-2\frac{a^2}{4}\sum_i \widetilde{x}_i^2\widetilde{y}_i^2}{\widetilde{x}^2\widetilde{x}^2(\widetilde{x}^2+m_d^2)(\widetilde{y}^2+m_d^2)},
\ee
with the negative and the positive sign coming from the first and second diagram, respectively, so that their sum cancels exactly. 

The sum of diagrams proportional to $g_3^4$ reads:
\begin{align}
\frac{g_3^4\Nc^2}{16\pi^2}\Bigg( &-\left[ \frac{5}{8} \Sigma^2 + (\frac{1}{2}- \frac{4}{3\Nc^2})\pi\Sigma  - 4(\delta+\rho) + 2\kappa_1 - \kappa_4 \right] \nonumber\\ &+ 2\rho + \frac{1}{2}\left[ \log\frac{6}{a \bar{\mu}} + \zeta \right] \Bigg)+\cont+\MSb
\end{align}
It is noteworthy that the scale dependence from the diagrams containing only gauge interactions with a single adjoint scalar field cancels exactly in the renormalization. However, upon the inclusion of another adjoint scalar field this property is lost.
\end{itemize}

\end{itemize}


\section{Diagrams for operator renormalization}
In this section, we give the results for the vacuum diagrams that affect the renormalization of quadratic condensates present in the action. The diagrams needed for the quadratic condensates $\langle \Tr Z^\dagger Z\rangle_{\MSb}$, $\langle \Tr M^\dagger M \rangle_{\MSb} $, and $\langle \Tr Z^\dagger \Tr Z \rangle_{\MSb}$:
\begin{itemize}
\item One-loop: 
\begin{align}
\begin{array}{c}
\vspace{0mm}
\includegraphics[width = 30pt,height = 30pt]{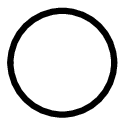}
\end{array}
&= \left\{ \begin{array}{ccl}
\phi & :  & \frac{1}{2}J(m) \\
\chi & :  & \frac{1}{2}J(m) \\
H    & :  & \frac{1}{2}(\Nc^2-1)J(m_d) \\
A    & :  & \frac{1}{2}(\Nc^2-1)J(m_d) \end{array}\right\} =(\Nc^2-1)J(m_d) + J(m)
\end{align}
\item Two-loop: the counter terms cancel the $\{c_3,d_3\}$-dependent linearly divergences terms, and only the gauge diagrams contribute:
\begin{align}
\begin{array}{c}
\includegraphics[width = 60pt,height = 30pt]{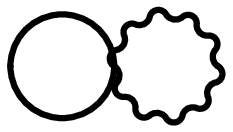}
\end{array}
&= \left\{ \begin{array}{ccl}
H    & :  & 1 \\
A    & :  & 1 \end{array}\right\} \frac{\Nc(\Nc^2-1)}{4}g_3^2\left[\frac{1}{a}I(0)-6I(0)I(m_d)-a^2m_d^2I(0)I(m_d)\right]
\end{align}
\begin{align}
\begin{array}{c}
\vspace{0mm}
\includegraphics[width = 30pt,height = 30pt]{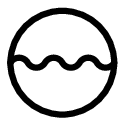}
\end{array}
= \left\{ \begin{array}{ccl}
A & :  & 1 \\
H & :  & 1  \end{array}\right\} \frac{\Nc(\Nc^2-1)}{4}g_3^2\big[ &-I(m_d)I(m_d)+4I(0)I(m_d)\nonumber\\
&-4m_d^2H(m_d,m_d,0)-a^2 G(m_d,m_d) \big]
\end{align}
\end{itemize}
The diagrams needed for the cubic condensates: $\langle 2\myRe\Det Z \rangle$ and $\langle 2 \myRe\Tr M^3 \rangle$:
\begin{itemize}
 \item Two-loop
\begin{align}
\begin{array}{c}
\vspace{0mm}
\includegraphics[width = 30pt,height = 30pt]{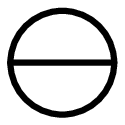}
\end{array}
= \left\{ \begin{array}{ccl}
\phi\phi\phi & :  & \frac{1}{18} \hat{c}_2^2\\
\phi\chi\chi & :  & \frac{1}{6} \hat{c}_2^2\\
\phi h h & : & \frac{1}{24}(\Nc^2-1)\hat{c}_2^2\\
\phi a a & : & \frac{1}{24}(\Nc^2-1)\hat{c}_2^2\\
\chi a h & : & \frac{1}{12}(\Nc^2-1)\hat{c}_2^2\\
hhh & : & 3(1/\Nc - \frac{5}{4}\Nc+\frac{1}{4}\Nc^3)(\frac{1}{3}\hat{c}_2+\hat{d}_2)^2 \\
aah & : & 9(1/\Nc - \frac{5}{4}\Nc+\frac{1}{4}\Nc^3)(\frac{1}{3}\hat{c}_2+\hat{d}_2)^2
 \end{array}\right\} g_3^6H(m_1,m_2,m_3)
\end{align}
\begin{align}
 =\left[ \left( \frac{\Nc^2-1}{6}+\frac{2}{9}\right) \hat{c}_2^2 + 12 \left(1/\Nc-\frac{5}{4}\Nc+\frac{1}{4}\Nc^3 \right)(\frac{1}{3}\hat{c}_2+\hat{d}_2)^2 \right]g_3^6 H(m_1,m_2,m_3)
\end{align}
\end{itemize}

\section{Basic lattice integrals and numerical constants}
In this appendix, we list the basic lattice integrals and numerical constants defined and calculated in \cite{Farakos:1994xh,Laine:1995ag,Laine:1997dy}.
\paragraph{Integrals:}
\begin{align}
 J(m)&\equiv\latint{x}\ln(\widetilde{x}^2+m^2)\nonumber\\
&=\frac{H(0)}{a^3}+\frac{1}{4\pi}\left[\frac{\Sigma m^2}{a}-\frac{2m^3}{3}+\mathcal{O}(a m^4)\right]\\
I(m)&\equiv\latint{x}\prop{x}{m^2}\nonumber\\
&=\frac{1}{4\pi}\left[\frac{\Sigma}{a}-m+\mathcal{O}(am^2)\right]\\
H(m_1,m_2,m_3)&\equiv \Latint{x}{y}\prop{x}{m_1^2}\prop{y}{m_2^2}\prop{x+y}{m_3^2}\nonumber\\
&=\frac{1}{16\pi^2}\left[\ln\frac{6}{a(m_1+m_2+m_3)}+\zeta+\frac{1}{2}+\mathcal{O}(am)\right]\\
G(m,m)&\equiv \Latint{x}{y}\frac{\sum_i \widetilde{x}_i^2 \widetilde{y}_i^2}{(\widetilde{x}^2+m^2)(\widetilde{y}^2+m^2)(\widetilde{x+y}^2)}\nonumber\\
&=\frac{1}{\pi^2}\left[\frac{\kappa_1}{a^4}-\frac{\delta}{4}\frac{m^2}{a^2}+\mathcal{O}(m^3a^{-1})\right]\\
H'(m_1,m_2,m_3)&=(\partial_{m_1^2})H(m_1,m_2,m_3)
\end{align}
\paragraph{Numerical constants:}
\begin{align}
 \Sigma&\equiv\frac{1}{\pi^2}\int_{-\pi/2}^{\pi/2}\dd^3x\frac{1}{\sum_i \sin^2(x_i)}\approx3.175911535625\\
 \delta&\equiv\frac{1}{2\pi^4}\int_{-\pi/2}^{\pi/2}\dd^3x\dd^3y\frac{\sum_i \sin^2(x_i)\sin^2(x_i+y_i)}{(\sum_i \sin^2(x_i))^2\sum_j \sin^2(y_j) \sum_k \sin(x_k+y_k)}\approx 1.942130(1)\\
\rho&\equiv\frac{1}{4\pi^4}\int_{-\pi/2}^{\pi/2}\dd^3x\dd^3y\left[\frac{\sum_i \sin^2(x_i)\sin^2(x_i+y_i)}{(\sum_i \sin^2(x_i))^2\sum_j \sin(x_j+y_j)}-\frac{\sum_i \sin^4(x)}{(\sum_i \sin^2(x_i))^2}\right]\frac{1}{(\sum_i \sin^2(y_i))^2}\nonumber\\
&\approx -0.313964(1)\label{rho}\\
\kappa_1 &\equiv \frac{1}{4\pi^4}\int_{-\pi/2}^{\pi/2}\dd^3x\dd^3y\frac{\sum_i \sin^2(x_i)\sin^2(x_i+y_i)}{\sum_i \sin^2(x_i)\sum_j \sin^2(y_j) \sum_k \sin(x_k+y_k)}\approx 0.958382(1)\\
\kappa_4&\equiv \frac{1}{\pi^4}\int_{-\pi/2}^{\pi/2}\dd^3x\dd^3y\frac{\sum_i \sin^2(x_i)\sin^2(x_i+y_i)\sin^2(y_i)}{(\sum_i \sin^2(x_i))^2\sum_j \sin^2(y_j) \sum_k \sin(x_k+y_k)}\approx 1.204295(1)\\
\zeta&=\lim_{z\rightarrow0}\left[\frac{1}{4\pi^4}\int_{-\pi/2}^{\pi/2}\dd^3x\dd^3y\frac{1}{(\sum_i \sin^2(x_i)+z)\sum_j\sin^2(y_j)\sum_k\sin^2(x_k+y_k)}-\ln\frac{3}{z}-\frac{1}{2}\right]\nonumber\\
&\approx 0.08849(1).
\end{align}

\bibliographystyle{tyyli}
\bibliography{articles.bib}

\end{document}